\documentclass[%
 amsmath,amssymb,
 aps, physrev,
]{revtex4-2}

\usepackage{graphicx}
\usepackage{dcolumn}
\usepackage{bm}

\begin{document}


\title{\textbf{Advancing Material Modeling in Hydrocodes Beyond Equations of State}
}%
\thanks{LLNL-JRNL-2009761}

\author{Tim A. Linke}
\email{Contact author: talinke@ucdavis.edu}
\affiliation{%
 Department of Mechanical and Aerospace Engineering, University of California, Davis, United States
}%
\affiliation{Materials Science Division, Lawrence Livermore National Laboratory, Livermore, United States}

\author{Dane M. Sterbentz}%
    \email{Contact author: sterbentz2@llnl.gov}
\affiliation{Materials Science Division, Lawrence Livermore National Laboratory, Livermore, United States}

\author{Jean-Pierre R. Delplanque}
\affiliation{%
 Department of Mechanical and Aerospace Engineering, University of California, Davis, United States
}%

\author{Sebastien Hamel}
\affiliation{Physics Division, Lawrence Livermore National Laboratory, Livermore, United States}

\author{Kevin A. Korner}
\affiliation{Materials Science Division, Lawrence Livermore National Laboratory, Livermore, United States}

\author{Philip C. Myint}
\affiliation{Physics Division, Lawrence Livermore National Laboratory, Livermore, United States}

\author{Lorin X. Benedict}
\affiliation{Physics Division, Lawrence Livermore National Laboratory, Livermore, United States}

\author{Jonathan L. Belof}
\affiliation{Materials Science Division, Lawrence Livermore National Laboratory, Livermore, United States}

\date{\today}

\begin{abstract}
We present a multiscale simulation framework that couples the Finite Element Method with molecular dynamics. Bypassing traditional equations of state (EOS) by using in-line atomistic simulations, the method offers the advantage of incorporating detailed microscale physics not easily represented with coarse-grained models. Coupling consistency with the continuum code is ensured through the use of lifting and restriction operators, in line with heterogeneous multiscale methods. The concurrent continuum-atomistic framework is validated through comparison with experimental results and conventional EOS models, and demonstrated in a shock-driven hydrodynamic flow simulation under extreme conditions. We further evaluate the framework's usability by comparing it to state-of-the-art EOS models of deuterium. A computational performance study reveals that the atomistic EOS evaluation is a feasible alternative to conventional approaches, and demonstrates a weak scaling of 99\% efficiency. These results highlight the framework’s potential for large-scale multiscale modeling across a broad range of materials and conditions.
\end{abstract}

\maketitle


\section{Introduction\label{sec:level1}}

Many phenomena in physics, engineering and material science exhibit behavior that span multiple length and time scales. Capturing such phenomena remains a central challenge in computational modeling. While continuum models offer computational efficiency, they inherently lack the formulation to account for microscopic insights. Conversely, fully atomistic simulations are still computationally intractable for systems of continuum scales. Multiscale modeling offers a promising compromise.

For instance, the accurate prediction of fluid flows requires a precise understanding of the behavior of their matter over a wide range of conditions. Particularly in the context of high-energy-density physics, the conditions within one simulation may span multiple orders of magnitude in temperature and pressure. Traditional methods for modeling such scenarios typically use hydrocodes (continuum-based solvers for highly dynamic problems) due to the large-scale nature of these flow fields. To capture the response of the underlying material, the continuum model relies on predefined equations of state (EOS). Typically, the equation of state is coupled to the conservation equations for mass, momentum, and energy, and ensures that changes in thermodynamic variables such as density, pressure, and internal energy are consistently reflected across these equations. For example, a widely used equation of state in hydrocodes is the Mie-Grüneisen EOS \cite{mie_zur_1903}\cite{gruneisen_theorie_1912}. It is formulated as
\begin{equation} \label{eq:MieGrun}
P = P_\textrm{ref}(V) + \Gamma\frac{E-E_\textrm{ref}(V)}{V} ,
\end{equation}
where $P_\textrm{ref} (V)$ is the reference pressure at a given volume $V$, $E_\textrm{ref} (V)$ is the reference internal energy at a given volume $V$, $E$ is the actual internal energy, and $\Gamma$ is the assumed constant Grüneisen parameter. Due to its analytical formulation, the Mie-Grüneisen equation does not require extensive computational resources when called. However, its limitations can emerge when encountering a wide range of conditions, beyond its range of validity (when e.g. assuming a constant $\Gamma$). Current state-of-the-art EOS designs counteract this by using the experimental and/or simulation results to choose functional forms for phase-dependent free energies \cite{wu_wide-ranged_2023}. This allows for highly accurate multiphase EOS in a wider range, but still cannot guarantee accuracy when materials deviate significantly from the conditions under which their EOS was created. Current equations of state can therefore become a significant source of error, especially in cases where the material behavior cannot be defined in advance. For example, porous materials can generate local hotspots under compression \cite{huy2023hugoniot}, or materials can undergo complex chemical reactions, making accurate free energy models challenging to construct \cite{wu2019equation}. Such phenomena most commonly occur when dealing with very high pressures and temperatures (e.g. GPa and $10^6$ Kelvin), and have become increasingly important for further advances in fields such as inertial confinement fusion \cite{haynes2016addressing}\cite{zylstra_burning_2022}. Since current equation of state modeling is subject to these fundamental limits, alternative approaches for simulating materials under extreme conditions are desired. 

One such approach is multiscale modeling, which integrates microscopic and macroscopic models in a coherent framework. By coupling an atomistic simulation to the continuum code, a multiscale approach can help overcome many of the constraints of its analytical or functional EOS counterparts. In particular, the \textit{equation-free} method offers a flexible framework to bridge the gap between microscopic dynamics and macroscopic behavior without relying on explicit macroscopic equations \cite{gear_equation-free_2003, divahar_two_2023, liu_acceleration_2015, frederix_equationfree_2007}. By leveraging microscopic simulations, such as molecular dynamics or kinetic Monte Carlo methods, the equation-free approach informs the macroscopic state of a system. Thus, it bypasses the challenging or impractical derivation of a constitutive relation (EOS and/or material strength model). The method is defined through iteratively lifting and restricting the system. The lifting operator translates macroscopic flow variables to microscopic states, and the restricting operator summarizes the microscopic results to update the macroscopic variables. Between the two operators, the system is evolved in the microscopic domain. Using these operators, a variety of sub-models can be assembled into a multiscale model \cite{knap_computational_2016}.

Therefore, the equation-free approach is not bound by the limitations of a traditional equation of state since the responsibility to accurately model the material behavior is passed onto the microscopic dynamics. Particle methods are essential in this context because they provide atomic-level interactions. A molecular perspective has proven essential when encountering complex phenomena such as phase transitions \cite{belof2023atomistic}. For instance, statistical atomistic methods such as those based in Classical Nucleation Theory (CNT) have been successfully used for phase transition modeling in continuum solvers \cite{myint2018nanosecond, myint2020coupling, sterbentz2019numerical, sterbentz2022model}. Various other methods like density functional theory (DFT) and quantum Monte Carlo (QMC) can replicate material properties in even greater, quantum-level detail. While these methods generally offer very high accuracies for EOS calculations, their computational costs are still prohibitive for the concurrent coupling approach presented in this work, where an independent EOS calculation is launched at every quadrature point in the hydrocode. DFT and QMC's vast computational requirements and CNT's overly simplistic assumptions call for a method that strikes a balance between them.

Classical molecular dynamics is an excellent candidate for the microscopic solver. These simulations can efficiently handle larger systems, longer timescales and almost arbitrary configurations, a necessity when bridging to macroscopic domains. They have also been shown to accurately capture complex phenomena under compression, such as phase transitions far from equilibrium \cite{sadigh_metastablesolid_2021} or hotspot formations \cite{huy2023hugoniot}, even when using relatively simple interatomic potentials. For cases requiring more sophisticated potentials, such as better agreement with DFT-based insights, classical molecular dynamics can be easily expanded \cite{zuo_performance_2020, willman2024accuracy}. For the remainder of this work, we will refer to classical molecular dynamics as simply molecular dynamics (MD). By leveraging the strengths of MD, the equation-free approach ensures that the macroscopic predictions are based on atomistic interactions.

While conceptually appealing, practical implementations of multiscale frameworks often face large computational costs. Equation-free methods are especially demanding, as they require many full microscopic solver evaluations in addition to the macroscopic model. As a response, new architectural and algorithmic enhancements continue to be developed \cite{alowayyed2017multiscale}. Machine learning techniques have shown successes in accelerating the concurrent evaluations \cite{leiter2018accelerated}\cite{lubbers2020modeling}. A computational breakthrough was provided by adaptive sampling and active learning methods that demand new microscopic insights based on an on-the-fly constructed database \cite{germann2020adaptive}\cite{karra_predictive_2023}.

Despite such successes, the approach has yet to find large-scale usage in engineering designs and has remained very domain-specific. Many factors contribute to the lack of adoption, including the complexity of mastering two or more distinct computational methods, the lack of a clear comparison to the performance of alternatives, and (still) the computational costs \cite{groen2019mastering}.

In this paper, we present conclusive arguments that multiscale simulations involving continuum methods at the higher scale and molecular dynamics at the lowest scale are feasible and effective for large-scale usage. We demonstrate this through the use of MD to capture material responses and inform a hydrodynamic simulation of their influence. We introduce a continuum-atomistic framework that concurrently couples macroscopic and microscopic simulations. Based on the equation-free approach, we bypass the formulation of an equation of state in hydrocodes. To achieve this, we introduce a series of operators to translate between the macroscopic and microscopic domain. Additionally, we develop a thermodynamic relaxation process for evaluating microscopic states from macroscopic properties. We employ MD simulations to capture the atomistic behavior of the underlying material, and use their predictions to update our macroscopic simulation. We deliver an implementation using two open-source solvers, described in detail in Section \ref{sec:eq-free}. The usability of the framework is shown for a problem involving a metal subjected to dynamic compression in Section \ref{sec:extreme}. Its performance for complex materials is discussed in Section \ref{sec:D2}. A key contribution of this work is the direct comparison between the proposed concurrent multiscale coupling and conventional approaches, analyzed with respect to physical accuracy as well as computational performance in Section \ref{sec:compute}. These results support the broader use of atomistically informed models within large-scale simulations as an alternative to purely continuum EOS-based simulations.

\section{Numerical Models}
\subsection{The Macroscopic Model}
For a complete formulation of flow fields, continuum descriptions such as the Euler equations are used. The Euler equations present the inviscid flow of a fluid for any general motion: 
\begin{align} 
    \label{eq:EulerMass} \frac{1}{\rho}\frac{\textrm{d}\rho}{\textrm{d}t} &= -\Vec{\nabla} \cdot \Vec{v} \\
    \label{eq:EulerMom} \rho \frac{\textrm{d}\Vec{v}}{\textrm{d}t} &= -\Vec{\nabla} p \\
    \label{eq:EulerE} \rho \frac{\textrm{d}e}{\textrm{d}t} &= -p\Vec{\nabla} \cdot \Vec{v}.
\end{align}
They define mass conservation (Eq. \ref{eq:EulerMass}), momentum conservation (Eq. \ref{eq:EulerMom}) and energy conservation (Eq. \ref{eq:EulerE}) using density $\rho$, the velocity field $\Vec{v}$, pressure $p$ and internal energy $e$ with respect to time $t$. To resolve the underdetermined system of three equations and four unknowns, a closure relation is required. This fourth equation relates the change of thermodynamic properties with respect to changes in the system:
\begin{equation}
    \label{eq:EOS} p= f(\rho, e) .
\end{equation}
Equation \ref{eq:EOS} defines the equation of state for the underlying material(s) of the model. Determining an accurate equation of state for a given material can be challenging due to complex and diverse effects under varying thermodynamic conditions. 

The Euler equations provide a great case study for this research as they find broad applications, including shock physics, vortex modeling and fluid instabilities \cite{dobrev_high-order_2012}. For this work, we use a solver based on the LAGrangian Higher-Order Solver (LAGHOS) \cite{dobrev_high-order_2012}. Laghos is an open-source solver developed and maintained by the Lawrence Livermore National Laboratory that uses the finite element method to discretize the time-dependent, compressible Euler equations in a moving Lagrangian frame using unstructured elements and explicit energy-conserving time-stepping \cite{code_laghos}. To enable shock capturing and improve the explicit time integration, we use the artificial viscosity model given in \cite{dobrev_high-order_2012}\cite{wilkins1980use}.

\subsection{The Microscopic Model}
In contrast to continuum mechanics, where partial differential equations are solved, atomistic simulations such as molecular dynamics track individual atoms or molecules and their interactions for a deterministic evolution of the flow field. This improves upon the gradient-based approach of partial differential equations since it inherently allows for discrete changes in the flow field, such as abrupt density fluctuations, that are difficult to represent using smooth, continuous variables. While a wealth of particle methods exist, classical MD most accurately captures material behavior on the large scale required to obtain macroscopic variable insights.

Molecular dynamics solves Newton's equations of motion for a system of interacting particles. In MD simulations, the positions and velocities of atoms and molecules are updated over time by numerically integrating these equations, e.g. by using the BBK \cite{brunger1984stochastic} or GJF \cite{gronbech2020complete}  integrators. The forces that act on the particles result from interatomic potentials and external force fields. The simulation starts by initializing the positions and velocities of all particles. The initial configuration most often does not correspond to a physically accurate system, so an equilibration step typically follows to allow the system to find a true physical state. The force between each particle is then evaluated, boundary conditions are considered, and positions and velocities are updated. The update occurs using numerical integration methods that keep the states within predetermined statistical ensembles. Various properties like temperature, pressure, and energy are obtained throughout the simulation by averaging the position and velocity of each particle \cite{wen_molecular_2022}. For this work, we use the Large-scale Atomic/Molecular Massively Parallel Simulator (LAMMPS) \cite{thompson_lammps_2022}, an open-source MD tool developed at Sandia National Laboratories. While it inherently captures the time and length scales of atomistic processes, computational advances have led molecular dynamics to bridge into experimentally observable time and length scales \cite{nguyen2021billion}. MD thus provides a dynamic picture of molecular systems and enables the observation of diffusion, phase transitions, and other macroscopic properties in real time. However, state-of-the-art molecular dynamics are only reaching nanoseconds and micrometer scales and are still far from continuum scales in both time and space. A coupled approach promises to maintain atomistic detail while simulating flow fields at the scales typically needed by engineering applications.

\subsection{FEM-MD Coupling}
\begin{figure}
\centering
\includegraphics[width=0.9\textwidth]{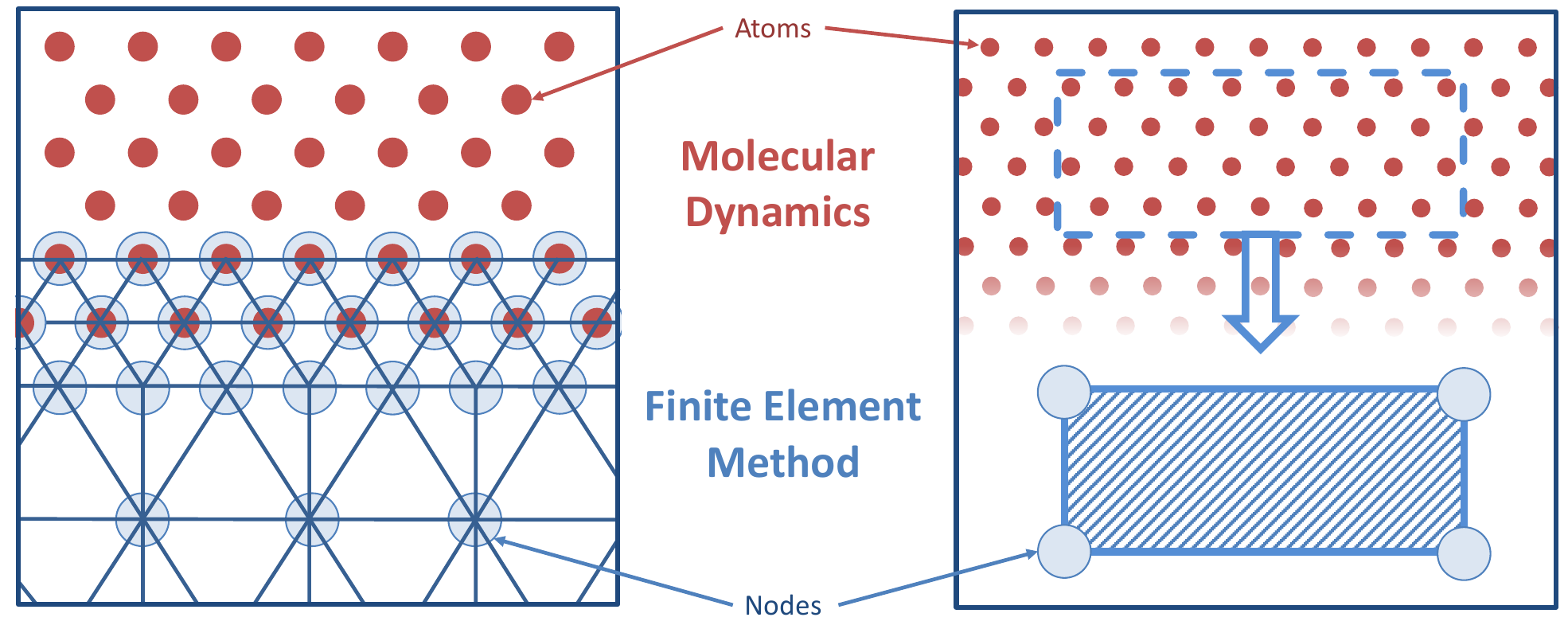}
\caption{Two main FEM-MD coupling approaches exist. The direct approach uses a handshake region (left) that overlaps atomic particles and continuum nodes to pass information, while indirect approaches (right) use support simulations.}
\label{fig:FEM-MD-methods}
\end{figure}

To achieve a coupled FEM-MD approach, many concurrent approaches have been developed. An excellent summary is provided by Lee \& Basaran \cite{lee_multiscale_2013} and Miller \& Tadmor \cite{miller2009unified}. Generally, these can be categorized into two main methods, shown in Figure \ref{fig:FEM-MD-methods}. The direct coupling approach combines MD and the finite element method by creating an overlapping region between the two domains, often referred to as a handshake region. Within this region, MD particles and FEM nodes exactly overlap. This overlapping area typically spans a distance equal to the cutoff distance of the interaction potential used in the MD region, ensuring that all particles in the transition zone have a complete set of neighbors within their interaction range. The handshake region ensures consistent forces between both domains by considering the interaction with their neighbors through the MD interaction potential while also calculating nodal forces in the FEM mesh. Forces and displacements are thus consistently transferred between the MD and FEM regions. To conserve mass within the region, the mass of each node is set to the mass of the corresponding MD particle. Additionally, both the MD and FEM regions typically use the same numerical integration scheme. \cite{smirnova_combined_1999}

The advantage of a direct physical continuum-atomistic coupling in the handshake approach has proven successful in a variety of applications \cite{lee_multiscale_2013, xiao2004bridging, smirnova_combined_1999, ganzenmuller_consistent_2012, kohlhoff_crack_1991, mao_atomistic-continuum_2015, rudd1998coarse, wagner2002coupling, delgado2003continuum}. Its implementation is often straightforward, as the region is predefined in space. However, it also introduces significant limitations. A fixed region where the coupling occurs requires prior knowledge of areas of physical importance, which may not always be available. In addition, the overlap between particles and nodes constrains the FEM simulation to MD length and time scales. To assure numerical stability in FEM, hydrodynamic scales can only be recovered far from the handshake region. Furthermore, forces must be evaluated twice for both the MD domain as well as the FEM mesh for each point in space, introducing significant computational costs.

The indirect (or hierarchical) coupling approach provides a solution to many of these issues. An indirect coupling selects one of the methods as the main descriptor of the model and supplements data using other models. The case of the atomistic model as the main solver is generally referred to as the atomistic finite element method (AFEM) or the quasicontinuum method, and is depicted on the right side of Figure \ref{fig:FEM-MD-methods}. It uses MD as the driver code and employs the finite element method as a support simulation. By continually running an MD simulation, the AFEM captures atomic-level processes very well. It dramatically improves the accuracy of continuum FEM, as it accounts for multi-body interactions. For cases where such detail is not required, it bridges into the continuum FEM domain through the translation of particles into nodes \cite{liu_atomic-scale_2005}. This has the advantage of summarizing the actions of many particles into fewer nodes, which speeds up the computation time significantly. It also eliminates artificial interface errors often encountered in the handshake region. The coupling can also occur at any temporal and spatial point throughout the domain, removing a critical restriction of the handshake region. The essence of this approach lies in the indirect connection between the two domains, and many cases have shown the success of the indirect approach \cite{liu_atomic-scale_2005, nasdala_molecular_2010, wu_atomistic_2006, shenoy1999adaptive}. However, given that MD is used as the driver code, it still limits applications to near-atomistic scales, especially temporal scales. 

To bridge the atomistic and continuum scales using an indirect method, the finite element method can be selected as the principal simulation tool of the coupled framework. By supplementing atomistic data from an MD simulation to the continuum solver, time and space remain independently coupled. In such frameworks, the continuum method requests atomistic insights in-situ. This class of methods is generally referred to as the Heterogeneous Multiscale Method (HMM). Specifically, we describe HMM problems of type B according to Ref. \cite{weinan2007heterogeneous}, for which a closed macroscopic model should exist, but lacks explicit definition to be used directly. The microscopic model serves to provide constitutive information and complete the macroscopic model.

HMM problems decouple length and time scales between the macroscopic and microscopic domain effectively, thus preserving fidelity and computational efficiency. This has led to HMM's adoption across many disciplines. In particular, material science benefits from accurate crystal plasticity mechanisms to inform constitutive behavior \cite{yamazaki2024multiscale, murashima2019coupling, nakane2000microscopic}. Similarly, the modeling of energetic materials has leveraged HMM to close complex thermodynamic relationships \cite{leiter2023temporal, barnes2019toward, barnes_lammps_2017}. In the broader context of fluid dynamics, HMM has seen applications in turbulence and interface dynamics \cite{smith2024multiscale}. It has also been applied to Lagrangian formulations \cite{moreno2023generalized}, which further generalizes the framework. These advances make the HMM especially suitable for this work.

\section{Equation-Free Equation of State Coupling} \label{sec:eq-free}
\begin{figure}
\centering
\includegraphics[width=0.8\textwidth]{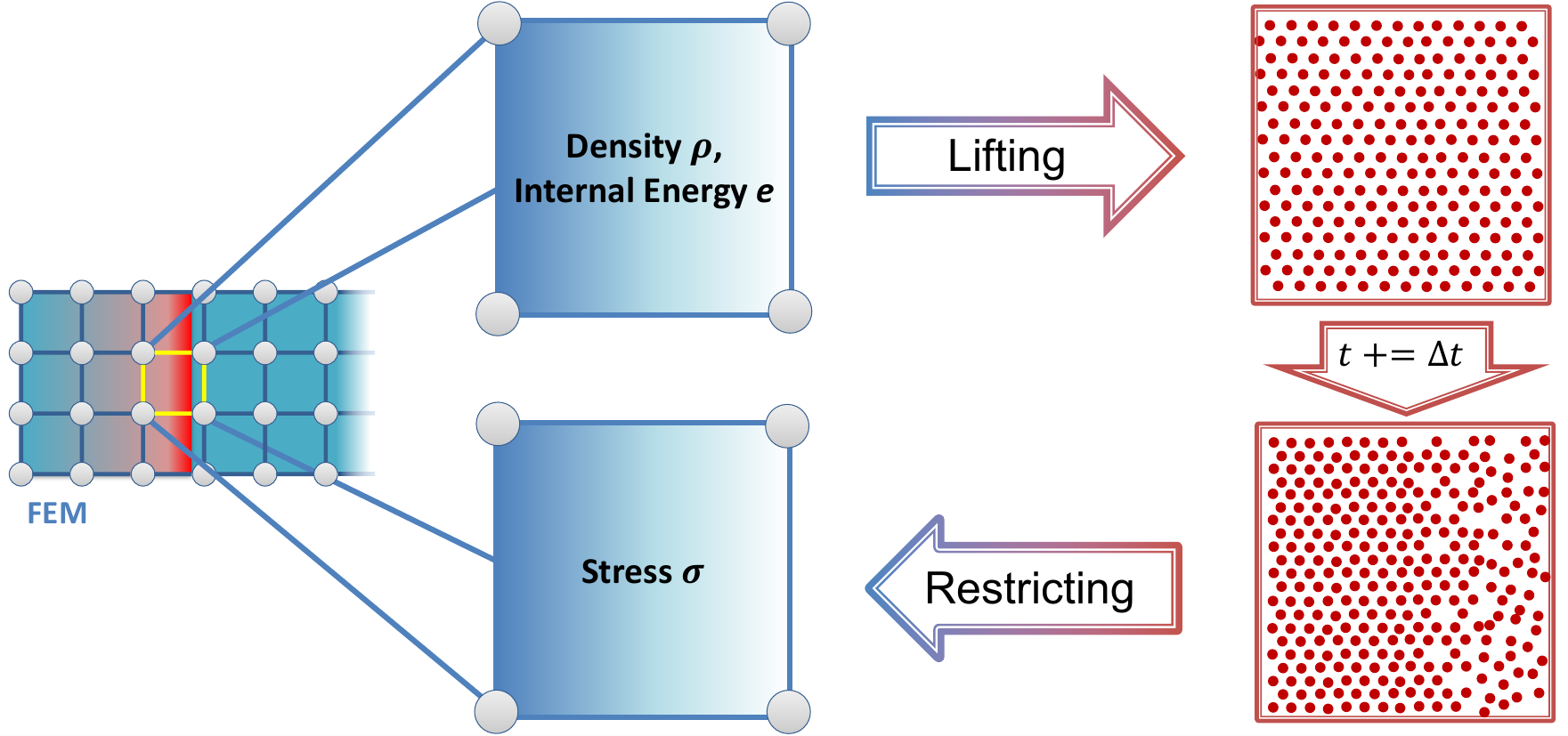}
\caption{The equation-free EOS coupling uses the lifting operator to initialize a microscopic instance and the restricting operator to inform the macroscopic domain of material properties.}
\label{fig:Eqfree}
\end{figure}

We present a computational framework that couples the molecular dynamics method to the finite element method following the indirect, HMM, equation-free approach. FEM serves as the primary solver for the macroscopic continuum equations. At each node in the FEM mesh, local quantities such as density, velocity and internal energy are evaluated. However, a relationship for each material used in the FEM mesh is necessary for determining thermodynamic properties. This thermodynamic relationship can be evaluated using a material equation of state, which is called at each quadrature point in the FEM domain and updated at each time step of the FEM simulation. The results are used to update material properties, such as stress, in the FEM domain. Employing the equation-free coupling to obtain the relationship among the thermodynamic properties, we design an alternative to conventional EOS approaches. When atomic-level detail is desired by the continuum method, MD simulations are invoked to update material properties. Figure \ref{fig:Eqfree} displays the main workflow of the approach.

The coupling framework is comprised of two principal operators. The lifting operator translates macroscopic variables, density and internal energy, into microscopic counterparts: the number density and atom velocities. It is responsible for initializing the MD simulation with an atomic configuration that accurately reflects the macroscopic state. To accomplish this, the lifting operator first considers the macroscopic density. Using the relation $n = \frac{N_A}{M} \rho$ between the molar mass $M$ of the substance, Avogadro’s constant $N_A$ and the macroscopic density $\rho$, we obtain the microscopic number density $n$. Then, an initial lattice structure of the given material must be selected as an initial state, and the number of atoms per unit cell $\alpha$ noted. Given the relationship between number density and the lattice spacing $l = \frac{1}{\sqrt[3]{n}}$, we initialize our system according to 
\begin{equation} \label{eq:numdens}
l = \sqrt[3]{\frac{\alpha M}{N_A \rho}}.
\end{equation}
The lattice parameter $l$ then initializes the position of all atoms within the MD domain. Through this translation, the lattice parameter makes the total number of atoms in the domain dependent on the volume of the system through the following relation:
\begin{equation} \label{eq:numAtoms}
N = \frac{V}{\alpha l^3}.
\end{equation}
This approach can represent both solid as well as liquid phases, as a liquid initialized on a lattice will naturally relax into a disordered state during equilibration.

Next, the macroscopic internal energy $e$ requires a microscopic interpretation. In almost all cases, the internal energy $e_\textrm{Hydro}$ in a continuum simulation equals the total energy in a microscopic system \cite{allen_computer_2017}:
\begin{equation} \label{eq:Econv}
    e_\textrm{Hydro} = E_{\textrm{kin,MD}} + E_{\textrm{pot,MD}}.
\end{equation}
The potential energy $E_{\textrm{pot,MD}}$ describes the contribution of atom positions and the resulting interatomic forces to the total energy, while the kinetic energy $E_{\textrm{kin,MD}}$ represents the contribution of atomic motion.

It is thus the objective of the lifting operator to split the macroscopic internal energy into its atomistic counterparts. With positions initialized, the system already occupies a configurational energy state. By initializing near-zero velocities, the initial potential energy can be found, which we denote as $E_\textrm{cold}$. Then, we begin an iterative process of approaching the equilibrated distribution among energies described in Equation \ref{eq:Econv}. First, we assume that all energy in the system is supplied by the kinetic energy. We convert the macroscopic internal energy to the temperature
\begin{equation}
    T = \frac{2e_\textrm{Hydro}}{N_\textrm{DOF}k_B},
\end{equation}
where $N_\textrm{DOF}$ is the number of degrees of freedom in the system and $k_B$ is the Boltzmann constant. Atom velocities are then initialized by drawing from the resulting temperature-dependent Gaussian distribution. Through an equilibration step, the system settles into its new state, and we sample the new potential energy $E_\textrm{pot}$. The new contribution required by the kinetic energy is obtained using the temperature
\begin{equation} \label{eq:Tcontribution}
    T = \frac{2(e_\textrm{Hydro}-(E_{\textrm{pot}}-E_{\textrm{cold}}))}{N_\textrm{DOF}k_B}.
\end{equation}
Equation \ref{eq:Tcontribution} highlights the necessity for $E_{\textrm{cold}}$; in a continuum model, the energy has a lower bound of $0.0$, while in a molecular dynamics context the total energy may be $E_{\textrm{tot}}<0.0$. $E_{\textrm{cold}}$ finds the true lower bound of a system and incorporates it into the energy distribution of Eq. \ref{eq:Econv}. $E_{\textrm{cold}}$ is dependent on EOS inputs $\rho$ and $e_\textrm{Hydro}$ and must therefore be calculated for every new MD system.

With the new $T$ of Eq. \ref{eq:Tcontribution}, the velocities are scaled to the corresponding distribution. The MD system is equilibrated again, the potential energy sampled and a new kinetic energy (temperature) found. We find that after two temperature update iterations, the system finds its physical distribution of $E_{\textrm{kin,MD}}$ and  $E_{\textrm{pot,MD}}$. With every atom’s position and velocity determined, the microscopic system is fully initialized. 

The state variables are then obtained in the canonical (NVT) ensemble. The automated nature of this approach presents the risk of unphysical states being propagated, and special care must be taken to ensure a convergence of results. Further details are provided in Appendices \ref{AppendixVal} \& \ref{AppendixOpt}.

Upon completion of the MD simulation, the restricting operator summarizes the detailed microscopic information back into macroscopic flow variables used in FEM. In the case of hydrodynamic simulations, the material reaction is reflected in the stress tensor. Equation \ref{eq:stress} shows that the macroscopic stress tensor of $[i,j]$ dimensions depends on the observed volume $V$, particle mass $m$, particle velocity $v$, particle position $r$ and particle force $f$ \cite{thompson_general_2009}:
\begin{equation} \label{eq:stress}
\sigma_{ij} = \frac{1}{V} \sum_{k=1}^{N} \left( m_k v_{ki} v_{kj} \right) + \frac{1}{V} \sum_{k=1}^{N'} \left( r_{ki} f_{kj} \right).
\end{equation}

Thus, the restricting operator is tasked with averaging the particle quantities and updating the stress tensor based on the atomic-scale computes. The pressure can be directly calculated from the stress tensor as the average of its normal components, and is coupled back into the continuum approach through Equations \ref{eq:EulerMom} and \ref{eq:EulerE}. This translation from microscopic to macroscopic states is the critical component of our approach, as it lifts any limitations on length scales that affected prior methods. Furthermore, the equation-free EOS coupling approach is not spatially or temporally restricted. At each node within the FEM mesh, the method makes real-time decisions whether to invoke MD simulations. This node-by-node evaluation ensures that MD simulations are used where atomic-scale information is necessary, which maintains computational efficiency.

As outlined, we implemented the concurrent framework in open-source software packages. In the microscopic domain, LAMMPS is employed. LAMMPS has a wide range of numerical integrators, interatomic potentials, and boundary conditions, and seamlessly scales from a single CPU to supercomputer architectures. This makes it particularly suitable for our approach.

In the continuum domain, Laghos is chosen as the driver code. Throughout every integration step, it calls a material model to obtain the stress tensor at current conditions. It is within this call that an instance of LAMMPS is created. The lifting operator initiates a microscopic simulation that runs concurrently, and the restricting operator returns the resulting stress tensor to Laghos. For all subsequent results, we launch an independent MD simulation at every quadrature point and every timestep. This intends to avoid potential inaccuracies introduced by approximation or acceleration strategies, and to demonstrate the full computational demands of the method.


\section{Application I - Extreme Conditions} \label{sec:extreme}
In conventional fluids simulations, the choice of equation of state is directly linked to the accuracy of the material response to the conditions within the flow. Often, the EOS is an analytical expression. No derivation is obtained without prior assumptions, and thus each analytical EOS inherits a range of validity. For instance, the commonly used ideal gas law assumes, among other simplifications, that atoms have negligible volume and interact through elastic collisions only. These assumptions become invalid when the material is exposed to low temperatures and/or high pressures, where interatomic forces and atomic volumes are relevant. 
A more sophisticated analytical EOS is the Mie-Grüneisen equation of state previously defined in Eq. \ref{eq:MieGrun}. It is commonly used for liquids and solids in high-pressure scenarios, such as shock wave compressions. It is defined through the following parameters: $P_\textrm{ref} (V)$ as the reference pressure,  $E_\textrm{ref} (V)$ as the reference internal energy at a given volume V, E as the actual internal energy, and $\Gamma$ as the Grüneisen parameter. It is therefore tailored to describe the deviation from a known reference state (often on the Hugoniot curve), which is particularly effective in material response modeling where the conditions vary significantly from standard conditions. It can even be designed to handle more complex behavior such as multiple phases by combining the EOS equations of each phase or multiple components \cite{hamel2012equation}. However, the Mie-Grüneisen EOS, as with any EOS model, is typically fit to pre-obtained data. This means it will always struggle when more complex, extreme behavior is present. \cite{colvin_extreme_2014}

To cover a wider range of conditions than analytical EOS models, one approach in high-energy-density science \cite{colvin_extreme_2014} is to model the EOS with a set of multiple EOS models, in which the model parameters are typically represented as splines, and write the resulting properties to tables defined over a discrete grid of temperature and density points. Interpolation mechanisms then bridge between data points to cover the phase space. This approach, often called tabulated equations of state, are known to capture complex phenomena that pure continuum EOS models may neglect \cite{foll2019use}. However, the tabulated EOS is also always restricted by the range of validity of its underlying data. Evaluation of states outside the range of validity requires extrapolation, which can lead to significant errors. As such, the encountered conditions within a simulation need to be known beforehand. This is not always possible, especially when dealing with highly dynamic problems at extreme conditions.

By obtaining equation of state data on-the-fly, the continuum-atomistic coupling allows for vast coverage of a material's phase space (provided that the underlying MD potential is suitably accurate). The effectiveness of this approach was first shown by \cite{ren_heterogeneous_2005}, and further adapted for highly parallel execution \cite{barnes_lammps_2017}. We note that the atomistic equation of state also has a range of validity: electronic effects such as ionization cannot be explicitly captured by classical molecular dynamics. Therefore, one must only apply such ion-only methods when the contribution from excited electrons can be neglected (e.g. when the Born-Oppenheimer approximation is valid). The responsibility of accuracy is passed down to the interatomic potential. 

In this section, we provide a direct comparison between an analytical, tabulated and atomistic equation of state. We consider a problem that is widely studied across the fluid dynamics community: shock propagation. A shock propagating in a material is an especially suitable example as the material undergoes large changes in density and internal energy, which can drive it to extreme conditions. The corresponding Hugoniot states are then a direct measure of the accuracy of the equation of state. To quantitatively assess the accuracy of the equations of state, we compare the results to experimental data.

\subsection{Simulation Setup}
A typical method to achieve Hugoniot states in a material is a flyer plate impact. An impactor is accelerated to high velocities and collides with the material to create a planar shock wave that propagates through the domain. A suitable material for the simulations is found in copper. Within the shock physics community, Cu is used as a pressure standard, and therefore has been widely studied \cite{sims_experimental_2022}. For this work, we draw upon a study by Mitchell et al. that provides shock compression data for EOS measurements of copper \cite{mitchell_shock_1981}.

In their work, Mitchell et al. \cite{mitchell_shock_1981} report on the impactor velocity $u_I$, impactor density $\rho_{0,I}$ and target density $\rho_0$. We initialized our domain accordingly. We tested two experimental cases to cover the low and high impact velocity domains. Table \ref{tab:init} summarizes the conditions.
\begin{table}[!h]
\centering
\renewcommand{\arraystretch}{1.2}  
\begin{tabular}{|c|c|c|}
\hline
$u_I$ \small{($\mathrm{km}/\mathrm{s}$)} &
$\rho_{0,I}$ \small{($\mathrm{g}/\mathrm{cm}^3$)} &
$\rho_0$ \small{($\mathrm{g}/\mathrm{cm}^3$)} \\
\hline
3.010 & 8.939 & 8.940 \\
6.913 & 8.948 & 8.936 \\
\hline
\end{tabular}
\caption{Initial conditions for the $V_\textrm{low}$ and $V_\textrm{high}$ compression cases.}
\label{tab:init}
\end{table}

\begin{figure}
\centering
\includegraphics[width=0.8\textwidth]{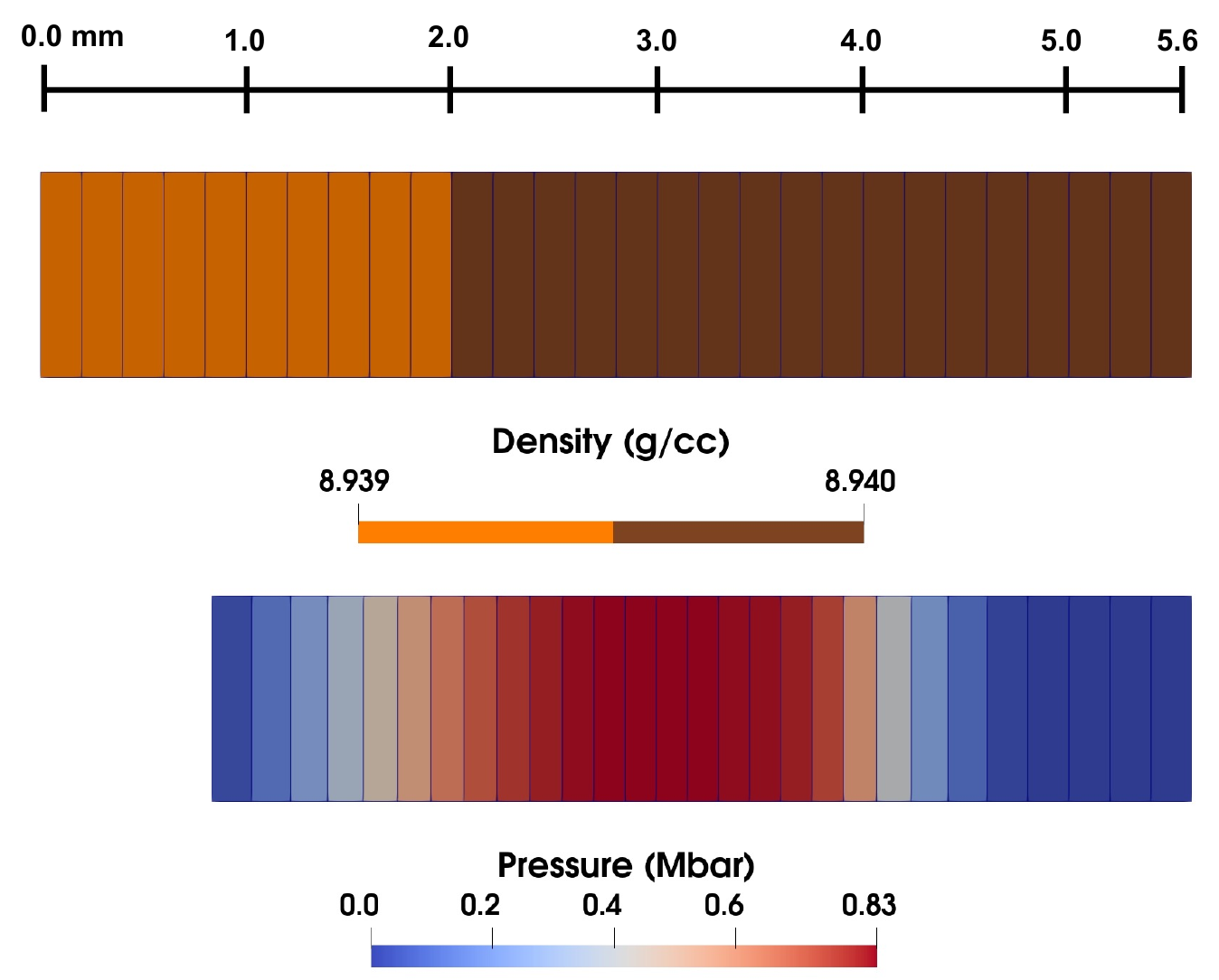}
\caption{The simulation domain of length 5.6mm is discretized using 28 elements and initialized to match experiments \cite{mitchell_shock_1981}. The density distribution (top) shows the impactor and sample at initialization. The pressure distribution (bottom) shows a post-impact compression state of the Lagrangian domain.}
\label{fig:simsetup}
\end{figure}

The experiment used $2$ mm thick impactors and measured the arrival of the shock in a plane positioned $1$ mm from the impact location. We match the experimental setup in our simulation by creating a 2D domain 5.6 mm in length and 1 mm in width. The simulation uses one element in width, making it quasi-1D. We find this assumption necessary, because the experiment reported on an axially symmetric distortion from planarity for the shock front. Using a sufficiently thin strip of material eliminates the requirement to model such curvatures. A depth of 5.6 mm was chosen to avoid any finite-domain effects that may occur once the shock front reaches the edge of the domain. We discretized the domain using 28 quadrilateral elements and set each simulation duration to allow ample time for the shock wave to traverse beyond $1$ mm of the sample (the measurement location of the experiments). Further details are given in Appendix \ref{AppendixOpt}. The simulation domain is shown in Figure \ref{fig:simsetup}. All simulations are run using a time step $\Delta t = 0.23$ ns. 

For our analysis, we compare the results of these simulations using three different EOS: an analytical Mie-Grüneisen EOS, a tabulated EOS, and our atomistic (concurrent MD) EOS. For the analytical equation of state, we employ a Mie-Grüneisen formulation with parameters matching the experimental results \cite{mitchell_shock_1981}: bulk sound speed $C=0.3933$ cm/$\mu$s, slope parameter $s=1.5$ and $\Gamma=2.12$. For the atomistic equation of state, we use the Mishin Cu interatomic potential \cite{mishin_structural_2001}, based on the embedded atom method. For the tabulated equation of state, we use a table directly derived from MD calculations using the Mishin Cu interatomic potential. This enables a direct comparison between the tabulated and atomistic equation of state. Details on the suitability of the tabulated EOS are given in Appendix \ref{AppendixSuit}.

\subsection{Low Velocity Impact}
Table \ref{tab:lowHug} presents the Hugoniot states at $x=3$ mm of the simulations using different EOS formulations for the low velocity experimental data with impactor velocity $u_I=3.010$ km/s. The experimental data provides the root mean square error \cite{mitchell_shock_1981}. While the analytical and tabulated EOS are deterministic, the atomistic EOS has inherent fluctuations. We present these deviations using the same root mean square error.
\begin{table}[h!]
\centering
\scriptsize
\renewcommand{\arraystretch}{1.2}  
\begin{tabular}{|c||c|c|c|}
\hline
$u_I=3.010$ (km/s) & Pressure (GPa) & Density (g/$\mathrm{cm}^3$) & Error in P / $\rho$ (\%) \\
\hline
Experimental Data & 83.71 $\pm$ 0.49 & 11.79 $\pm$ 0.03 & -- \\
Analytical EOS & 83.16 & 11.83 & 0.66 / 0.34\\
Tabulated EOS & 84.45 & 11.80 & 0.88 / 0.08\\
Atomistic EOS & 84.139 $\pm$ 0.180 & 11.819 $\pm$ 0.005 & 0.51 / 0.25 \\
\hline
\end{tabular}
\caption{Hugoniot states for the low velocity compression case}
\label{tab:lowHug}
\end{table}
We observe general agreement among all three equations of state with experimental data. The analytical EOS shows pressure deviations from experiments by a relative error of $0.66\%$, while the tabulated EOS has a relative error of $0.88\%$ and the atomistic EOS differs by $0.51\%$. The resulting density data shows similar agreements: the analytical EOS shows a relative error in density of $0.34\%$ and the tabulated and atomistic EOS show density deviations of $0.08\%$ and $0.25\%$, respectively. All three methods agree exceptionally well, confirming the validity of the simulation setup with respect to the experiment.

\subsection{High Velocity Impact}
The high velocity compression case of $u_I=6.913$ km/s presents a more significant challenge to the equations of state, as the material is compressed to larger pressures and densities. Nonetheless, all three EOS models should show good agreement with the experiment. Table \ref{tab:highHug} presents the Hugoniot states at $x=3$ mm of the simulations using different EOS formulations for the high velocity impact experiments. The experimental data and atomistic equation of state also provide the root mean square error.
\begin{table}[h!]
\centering
\scriptsize
\renewcommand{\arraystretch}{1.2}  
\begin{tabular}{|c||c|c|c|}
\hline
$u_I=6.913$ (km/s) & Pressure (GPa) & Density (g/$\mathrm{cm}^3$) & Error in P / $\rho$ (\%) \\
\hline
Experimental Data & 282.7 $\pm$ 2.9 & 14.36 $\pm$ 0.09 & -- \\
Analytical EOS & 269.85 & 14.27 & 4.55 / 0.63 \\
Tabulated EOS & 275.02 & 14.57 & 2.72 / 1.46 \\
Atomistic EOS & 276.195 $\pm$ 0.768 & 14.286 $\pm$ 0.001 & 2.30 / 0.52 \\
\hline
\end{tabular}
\caption{Hugoniot states for the high velocity compression case}
\label{tab:highHug}
\end{table}
Generally, excellent agreement is observed in the density measurements. The experimental value of $\rho = 14.36 \ \textrm{g}/\textrm{cm}^3$ is closely replicated by all three methods, with a maximum relative error of $1.5\%$ (tabulated EOS) among the three. Deviations in pressure between the experiment and the simulations are more pronounced. The relative error in pressure for the analytical EOS is the largest, showing $4.5\%$. The tabulated EOS improves to a relative error of $2.7\%$, and the atomistic EOS shows the best agreement with a relative error of $2.3\%$. Although the error in pressure increased compared to the low velocity impact case, a maximum relative error of $4.5\%$ still signifies very good agreement, validating the approach when handling higher pressures.

\subsection{Discussion}
The results of the low velocity impact show excellent agreement between experiment and any of the three computational models, solidifying the general description of the problem by the continuum code. In addition, it highlights that all three methods are capable of accurately capturing the material response at large compressions in the range of Gigapascals and $\frac{V_0}{V}\leq 0.75$ ($V$ is volume). We observe slightly larger deviations for the high compression case in both pressure and density. In addition, we observe that in contrast to the low velocity case, which showed the simulation predicting both higher and lower pressures, the simulations of the high velocity case consistently underpredict the pressure. 

The errors between computational and experimental data can result from a range of causes, all of which can be categorized as numerical or physical errors. With regards to the continuum solver, the numerical error is most pronounced during the discretization of the domain. While the low velocity impact case exhibits excellent agreement, especially in density, we observed a much shorter-lived duration of the Hugoniot state for the high velocity impact. As such, the discretization plays a key role. Sharp rises in pressure, density, and internal energy are observed immediately behind the shock wave. If the temporal discretization is too large relative to the duration of this transient state, the results average out the true peak. Similarly, a coarse spatial discretization may smear the steep pressure gradient. As a result, the simulated pressure profile will underpredict the true magnitude and sharpness of the Hugoniot state. A further analysis and discussion of the discretization choices is found in Appendix \ref{AppendixOpt}.

When discussing the physical error of the simulation, we initially observe that the match in density outperforms the agreements in pressure. Since density is prescribed as an input and pressure as an output of the equations of state, these deviations are likely to be a direct consequence of the chosen EOS. The potential errors of each EOS model are discussed here.

The Mie-Grüneisen EOS is based on the simple and (mostly) effective formulation that expresses pressure as a function of a reference volume and internal energy. Although its parameters can be tailored to fit high-compression cases very well, the underlying linear relationship between properties may deviate from material behavior in the experiments. In this study, parameters derived directly from the experimental data are used. This is reflected in the excellent agreement at low impact velocities, and with the high-velocity impact density result. The $4.6\%$ pressure error for the high velocity case therefore indicates that the material behavior at these conditions may deviate from the linear relationship assumed by the EOS.

The tabulated equation of state has two main sources of errors: the accuracy of its underlying data and an interpolation/extrapolation error. With regards to its underlying data, the tabulated EOS was derived from MD-based calculations using the Mishin Cu potential (specifically, the bcc and fcc melt curves predicted by that potential). It is therefore expected to yield very similar results compared to the atomistic EOS. While the low velocity case obtains very similar results between the two methods, the density measurement in the high velocity case stands out among the comparison (see Table \ref{tab:highHug}).
The interpolation error is investigated further. It can be minimized by both employing a fitting interpolation scheme as well as ensuring that the data to which it is fit covers a range which is appropriate for the intended applications. The suitability of the EOS model as well as its interpolation method has been ensured, a detailed analysis of which can be found in Appendix \ref{AppendixSuit}. Nonetheless, if the range of validity of the table is exceeded, the requested thermodynamic states are obtained using extrapolation. We found that the energy values passed to the tabulated EOS exceed its calibration range only for the high velocity impact. The range of validity of the table is not challenged during the low compression case, explaining the excellent agreement among the equations of state. The extrapolation of thermodynamic states in the table may decouple the relationship between density and internal energy, which explains the evident density mismatch on the continuum solver side.

The atomistic EOS relies on the reliability on the coupling framework as well as the physical accuracy of the MD simulations. The coupling framework's reliability, which had been extensively verified for this study, is also validated with the excellent agreement of the low velocity impact measurements. In contrast, the physical accuracy of the MD simulations are heavily reliant on the interatomic potential. The Mishin Cu potential was originally derived for low density cases \cite{mishin_structural_2001}, but had been shown to yield accurate results even at higher compression cases \cite{sims_experimental_2022}. Nonetheless, it is not guaranteed to work perfectly at these conditions, and discrepancies such as those seen in Table \ref{tab:highHug} likely arise from it. The interatomic potential is therefore a crucial choice for any atomistic EOS.

Lastly, the experimental data itself is reported with error bounds. For the density, the bounds of $\pm 0.2\%$ and $\pm 0.6\%$ for the low and high velocity impact case, respectively. For both cases, the values of the simulation results fall almost always entirely within these bounds. For the pressure, the bounds for the low and high velocity case are reported as $\pm 0.6\%$ and $\pm 1.0\%$, respectively. These bounds could also account for a significant portion of the percentage errors obtained by the simulations.

In conclusion, very good agreement among well established equations of state, such as analytical or tabulated approaches, is found with the atomistic framework presented in this work. We confirm the validity of the simulations with experimental data of weak and strong shock compression cases. In essence, two main errors are identified: the numerical (discretization) error and the physical accuracy. The thermodynamic states that are present during the strong compression challenge the calibration range of both the analytical and tabulated equation of state. For a case in which we were convinced of the accuracy of the interatomic potential over the full range of conditions, integration of the atomistic EOS into the continuum simulations, as made possible with our approach, would prove invaluable.

\section{Application II - Complex Materials} \label{sec:D2}
In the previous section, we investigated the behavior of copper at extreme conditions. We chose copper as it is a well studied material due to its relatively simple behavior across varying conditions. This means that existing EOS models, such as a Mie-Grüneisen formulation, are capable of representing copper well. The true advantage of running an atomistic EOS is for materials which are highly complex, exhibiting vastly different behavior within the exposed thermodynamic conditions. Here, analytical expressions are immediately limited and tabulated interpolation schemes become increasingly difficult to design. The in-line atomistic simulations can, in principle, obtain constitutive properties “on-the-fly”.

One such example of a complex material is deuterium, a stable isotope of hydrogen. It finds application, among others, in fusion energy research. Throughout an inertial confinement fusion (ICF) experiment, deuterium will experience temperature ranges from cryogenic to tens of millions of degrees and pressure peaks from near-vacuum to more than a billion atmospheres \cite{betti2016inertial}. The span across orders of magnitude makes the accurate prediction of deuterium's behavior especially difficult. At the same time, accurately modeling deuterium is critical: its equation of state is responsible for correctly capturing the compressibility of the fuel \cite{hu2010strong}, drive efficiency \cite{robey2012shock} and instability formations \cite{hammel2010high}, all of which have a direct influence on the total fusion yield. For instance, a decrease in ablation pressure of $10\%$ was shown to have significant consequences to the implosion process \cite{hu2015first}, making the equation of state an essential area of improvement in ICF modeling \cite{haynes2016addressing}.

To address this issue, a workshop was conducted in 2017 to compare state-of-the-art equations of state for ICF-relevant materials, including deuterium \cite{gaffney_review_2018}. The comparison entailed EOS constructed from first-principle simulations and experimental data, all of which were provided in tabular form. We revisit the workshop's effort by contributing the performance of our atomistic EOS. 

The comparison was made for thermodynamic relations along a Hugoniot of deuterium, starting from a cryogenic and compressed state. The widely used tabulated \textit{2003-Kerley} EOS \cite{kerley2003equations} is used as the benchmarking case. We note here that the original report refers to the \textit{2003-Kerley} benchmark as \textit{LEOS 1014}. From the \textit{2003-Kerley} EOS, we obtain the Hugoniot curve and derive the thermodynamic relations using our continuum-atomistic framework. We note that no major changes to the framework were required, with the exception of adding a new interatomic potential for deuterium. We used a Hydrogen machine learning potential trained on DFT data, presented in \cite{cheng_thermodynamics_2023}, and adjusted the mass to account for the additional neutron. The use of the hydrogen potential for deuterium is justified at the conditions of interest, as differences arising from nuclear mass become negligible and the electronic potential governs the system’s behavior.

\begin{figure}
\centering
\includegraphics[width=0.8\textwidth]{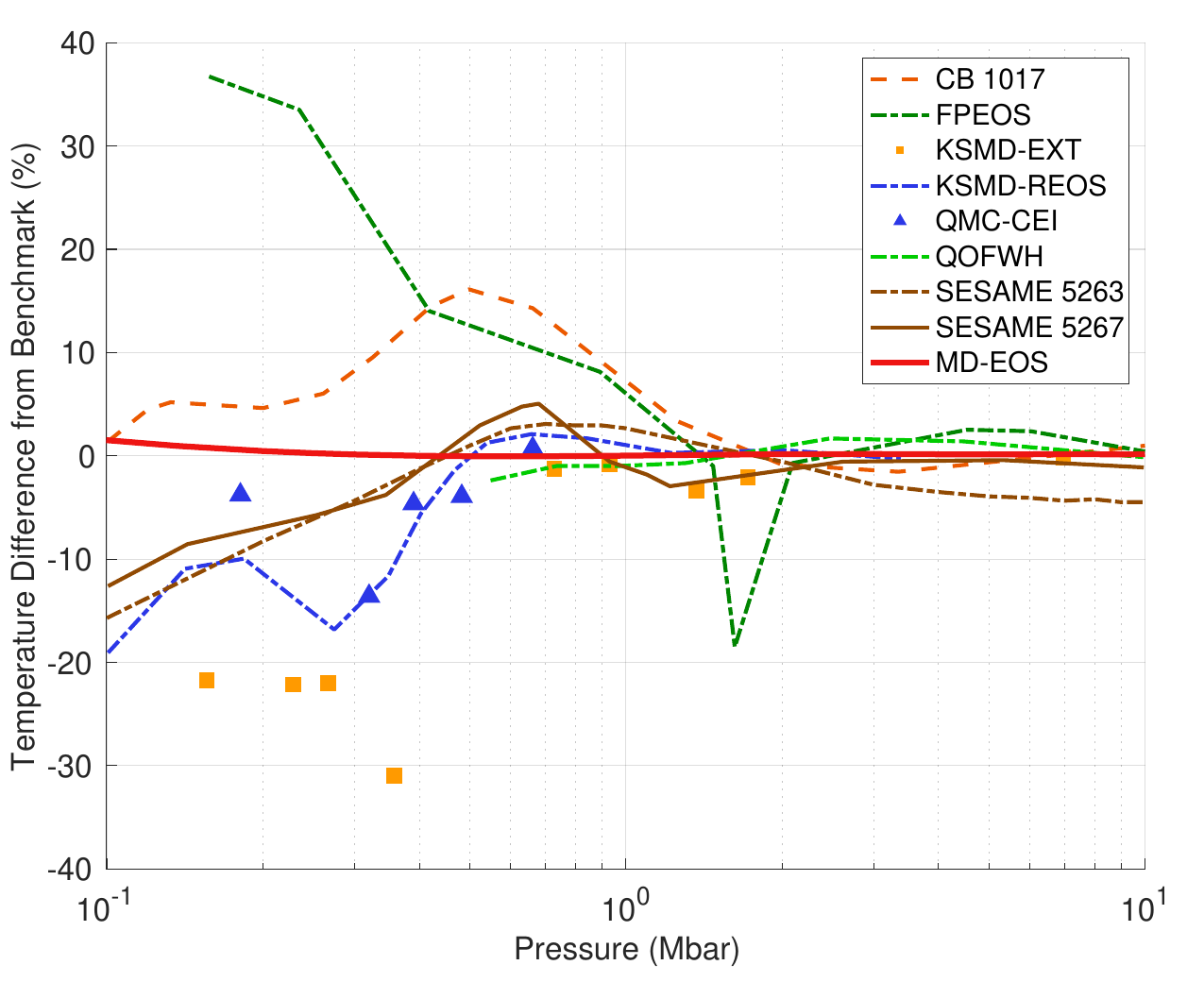}
\caption{A comparison between the atomistic EOS of this work (red) with state-of-the-art D$_2$ EOS models \cite{gaffney_review_2018} along a Hugoniot. Temperature differences are measured against the \textit{2003-Kerley} EOS model \cite{kerley2003equations}.}
\label{fig:D2study}
\end{figure}

Figure \ref{fig:D2study} shows the temperature difference of the equations of state at pressures ranging from 0.1 to 10 Mbar. The temperature-pressure relationship was obtained along a Hugoniot using initial conditions $\rho_0=0.171 \ \textrm{g}/\textrm{cm}^3$  and $T_0=20$ K \cite{gaffney_review_2018}. Overall, two main regions are identified: a low pressure region ranging from 0.1 to 1 Mbar, and a high pressure region from 1 to 10 Mbar. General agreement among most methods is observed in the high pressure region. The atomistic EOS, shown in red, agrees very well with the existing EOS models. These tabulated models are among the most accurate for inertial confinement fusion materials \cite{gaffney_review_2018}. They are based on high-fidelity computational models such quantum molecular dynamics and/or fits to experimental data. Our atomistic equation of state is based on classical molecular dynamics and therefore calculates properties with a lower fidelity than the compared models. Nonetheless, we observe excellent agreement with almost all methods in the high pressure region of the Hugoniot. This indicates that our approach, employing the MD potential of Ref. \cite{cheng_thermodynamics_2023}, can accurately capture the physics governing the material under extreme compression, such as many-body interactions and short-range repulsive forces. Consequently, the atomistic method offers an alternative for exploring high-pressure phenomena of deuterium. 

In the low pressure region (0.1 to 1 Mbar), large deviations among all models are observed. In this range, molecular dissociation and other quantum effects are relevant, and it is apparent that no method agrees on the corresponding temperatures. In this regime, mean-field descriptions are known to encounter difficulties due to the low particle density. Errors up to $40\%$ are observed among state-of-the-art methods. Agreement within the low pressure region of deuterium is still an active research field, and improvements to tabulated approaches continue to be made \cite{mihaylov2021improved}. Refinement of our classical molecular dynamics simulations in the low-pressure regime to match e.g. experimental data is subject to future work. Nonetheless, Fig. \ref{fig:D2study} shows that the atomistic EOS closely replicates the behavior of the \textit{2003-Kerley} EOS, showing a maximum relative error of $2\%$ (at $p=0.1$ Mbar). The close match with the \textit{2003-Kerley} EOS, the commonly used EOS in ICF-type simulations, indicates that the atomistic EOS can be considered as a candidate for future research.

\section{Computational Performance} \label{sec:compute}
To solidify the concurrent continuum-atomistic framework as a candidate for future design and research efforts, the computational performance is presented. In the EOS review report of 2018, the authors noted that "it is not computationally feasible to have large-scale simulation codes also calculate material properties like the EOS ‘on-the-fly’"\cite{gaffney_review_2018}. In the previous section, we revisited the workshop's effort by contributing our (classical) atomistic EOS to the comparison. In this section, we shine new light onto the feasibility of the approach.

\begin{figure}
\centering
\includegraphics[width=0.65\textwidth]{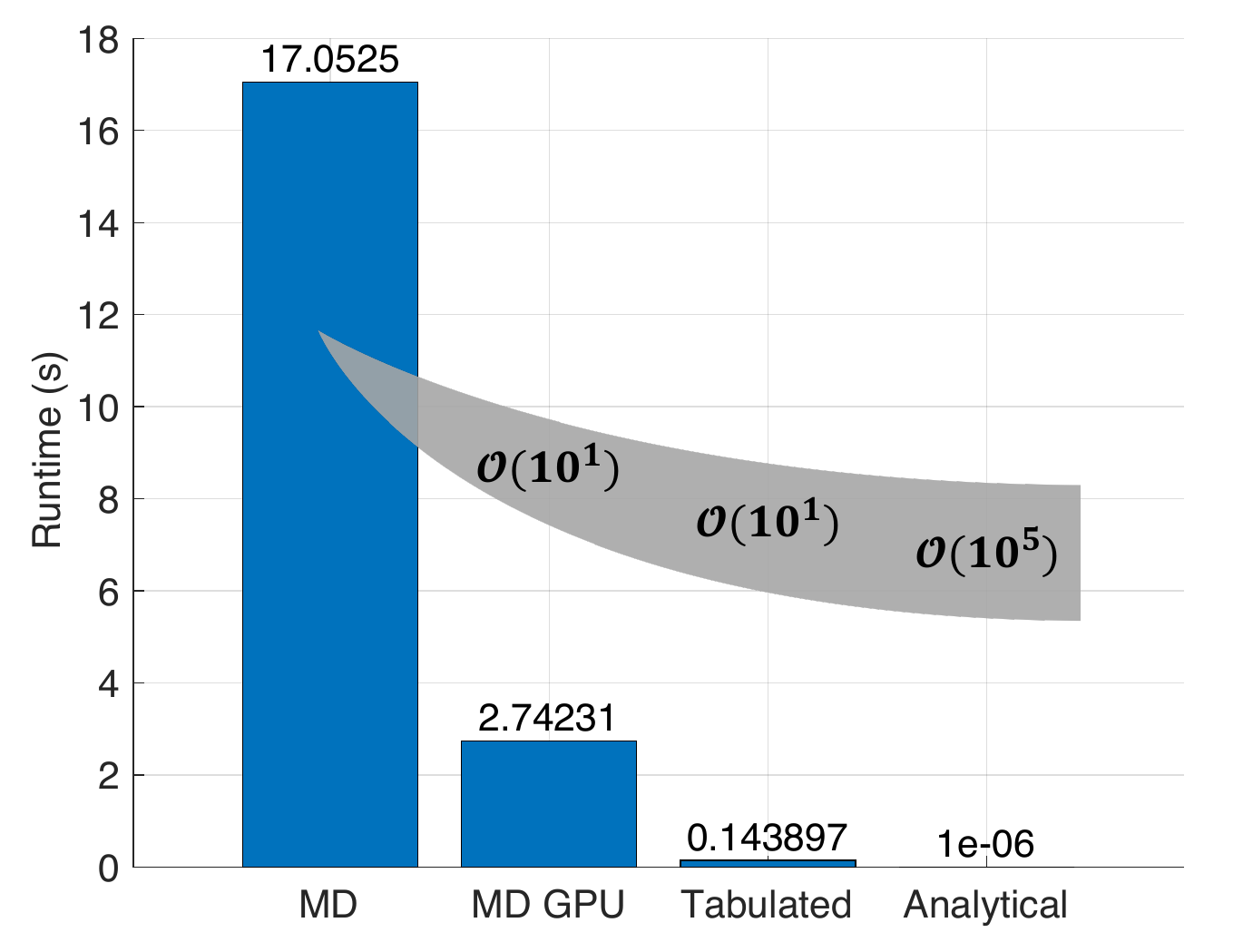}
\caption{A comparison between runtimes of different EOS models}
\label{fig:EOSSpeed}
\end{figure}

In a continuum simulation, the equation of state is called at every instance of the material model. In a simulation discretized using the Finite Element method, it occurs at every quadrature point. This makes it one of the most-called operations in the simulation, and computational efficiency is a key aspect to the overall performance of the simulation. Figure \ref{fig:EOSSpeed} presents the typical runtime of different choices in EOS modeling. The atomistic EOS times shown include the creation of a new LAMMPS instance every call, the MD solve and returning the EOS result. For the tabulated EOS, we measured the runtime our Mishin Cu table takes to load, read and return the table, and the analytical EOS is a function call to the Mie-Grüneisen model. Unsurprisingly, the analytical equation of state, which requires only a few arithmetic computations per evaluation, has the shortest runtime at $T=10^{-6}$ seconds. For tabulated equations of state, the runtime increases dramatically by five orders of magnitude to $T=0.144$ seconds. In the case of the atomistic EOS, we observe a very long runtime of $T=17.05$ seconds. In this case, the continuum code as well as the molecular dynamics simulations are run solely on CPUs. For such long runtimes, we must agree with previous assumptions that an 'on-the-fly' analysis is infeasible. However, molecular dynamics simulations are very parallelizable, especially on GPUs. When running the continuum code on CPUs, but passing the molecular dynamics simulations to GPUs, we observe runtimes of $T=2.74$ seconds. This improves the runtime of the atomistic EOS by a factor of more than 6, and places it within one order of magnitude of the tabulated equation of state. For problems whose materials are too complex or conditions are too extreme to be accurately captured in tabulated models, the atomistic equation of state presents a viable alternative. 

We note that the demands of the molecular dynamics system for this work were small, typically containing $\mathcal{O}(10^3)$ atoms (see Appendix \ref{AppendixOpt}). While such systems proved to capture the relevant phenomena in this work, future applications may require much larger systems and more timesteps. For such scenarios, we observed that larger system size as well as more timesteps increased the CPU-based MD calculation runtime dramatically. However, the GPU-based MD calculation runtimes remained within the order of magnitude of the presented result in Fig. \ref{fig:EOSSpeed}. Therefore, we strongly recommend the use of GPUs for the molecular dynamics simulations. When running concurrent continuum-atomistic simulations, the use of GPUs implies either a GPU-compatible continuum code or very frequent communications between CPUs (continuum) and GPUs (atomistic). GPU-enabled continuum codes promise less communications among the hardware, but face the difficulty of cumbersome implementations. Because continuum-atomistic frameworks often couple two standalone codes, such as the one presented in this work, seamless GPU usage requires the unraveling of external library calls, redefinitions of internal communications and more. In contrast, Fig. \ref{fig:EOSSpeed} indicates that the main computational costs will result from the MD calculation, and the speed up gained by using GPUs in the continuum code may be negligible. Therefore, optimizing the communication between CPUs and GPUs is important. Modern hardware architectures often integrate CPUs and GPUs on a single chip and provide access to shared memory. Known as Accelerated Processing Units (APUs), they find usage in today's most powerful supercomputers like El Capitan \cite{llnl_elcapitan}. Shared memory reduces data transfer latency and significantly streamlines communication. Such architectures are therefore particularly attractive for HMM simulations, so we use APUs for the concurrent continuum-atomistic framework presented in this work.

\begin{figure}
\centering
\includegraphics[width=0.9\textwidth]{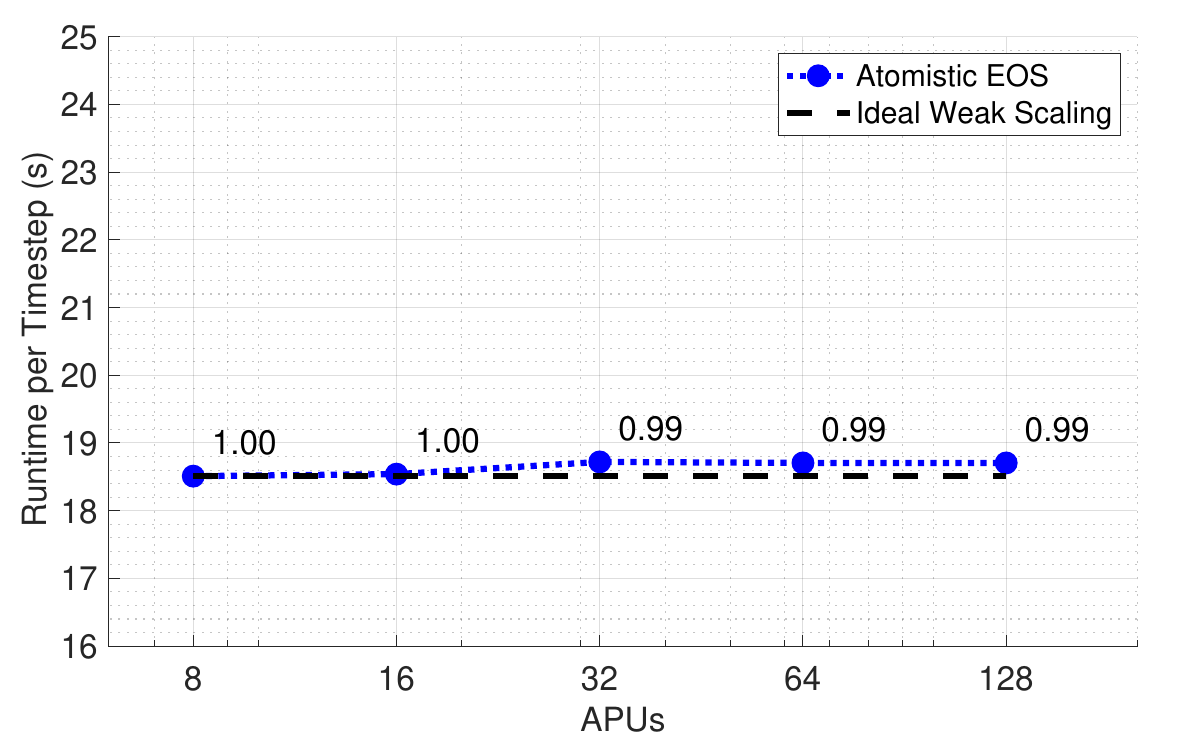}
\caption{Weak scalability results of our atomistic EOS, from 8 to 128 APUs}
\label{fig:scaling}
\end{figure}

In addition to assessing the computational efficiency of our framework, we address the suitability of the framework for large-scale simulations. The performance metric that primarily matters for most applications is the runtime per timestep. A suitable framework should maintain a consistent runtime as the total problem size increases proportionally with the number of computational resources. This ensures that problems are not restricted by bottlenecks in the implementation. In Fig. \ref{fig:scaling}, we report our weak scalability study. 

The scalability test was conducted on LLNL's Tuolumne system, which uses \textit{AMD\texttrademark \ MI300A} APUs that consist of \textit{CDNA 3} GPUs and \textit{4th Generation AMD EPYC} CPUs \cite{tuolumne}. The simulation set-up followed the low velocity impact test presented in Section \ref{sec:extreme}. For our scalability test simulations, reported runtimes were averaged over forty consecutive timesteps. We assigned each APU to one element of the domain, thereby maximizing the communication among APUs. The solver maintained 99\% weak parallel efficiency over a 16-fold increase in problem size. The excellent scaling behavior is an important step towards the large-scale usability of the concurrent continuum-atomistic framework.

\section{Conclusion}
We presented a framework for coupling the Finite Element Method with molecular dynamics, enabling simulations that bridge between microscopic and macroscopic scales. The approach bypasses traditional equations of state, relying on interatomic potentials to describe material behavior. A strategy for consistent coupling is introduced through the lifting and restricting operators. We validated the framework by comparing to experimental data as well as conventional EOS models. The coupling of hydrodynamics and MD was demonstrated using a shock-driven flow simulation at extreme conditions. We found that the atomistic, tabulated and analytical EOS perform well for the problem. We highlighted the usability of the framework when considering complex materials, and positioned the method alongside current state-of-the-art equations of state to assess its accuracy. Lastly, the computational performance of all three EOS models were directly compared, with the atomistic EOS evaluation landing within one order of magnitude in runtime compared to conventional approaches. We highlighted how exascale-computing architectures such as APUs are strong enablers of this work. A weak scalability study showed a $99\%$ efficiency of the framework over a 16-fold increase in problem size, supporting large-scale usage of the method.

While agreement with current methods paves the way for design purposes, the prospect of using the atomistic EOS goes beyond the pressure-temperature-density relationships of an equation of state. Currently, there is no consensus on how to handle the equation of state for continuously varying mixtures. In addition, equations of state for foam materials, which have shown promise for ICF \cite{olson2016first}, are difficult to represent properly using simple tabular models \cite{paddock2023measuring}. Lastly, non-equilibrium phenomena and time-dependence are highly relevant \cite{kabadi2021thermal}, but present major difficulties to tabulated equations of state. It can be, however, a necessity to guarantee accuracy in these regimes. Nonequilibrium effects are often responsible for crucial processes such as energy dissipation and entropy production in turbulence \cite{verma_boltzmann_2020}, or phase transitions \cite{belof2023atomistic}. In fact, nonequilibrium processes have been shown to lead to kinetic stabilization of metastable phases \cite{sadigh_metastablesolid_2021}. These metastable phases can significantly influence transport coefficients in continuum mechanics. Understanding nonequilibrium effects is therefore essential to advance our description of the behavior of matter. Therefore, and perhaps most importantly, we note that current EOS models used in radiation continuum codes lack self-consistency with other physics models, such as transport properties or strength models\cite{haynes2016addressing}. The framework presented here has the fundamental ability to resolve all of the listed issues, as it can provide information about the atomic system above and beyond the purview of bulk, thermodynamic properties. Future investigations should therefore target a more sophisticated integration of the microscopic model into the continuum model, such that additional phenomena can be described consistently across scales using the same microscopic description. 

Connecting transport properties, strength models, and equations of state to circumvent the need to prescribe a material model would mark a major step toward resolving one of computational physics’ fundamental challenges, and this work introduces a viable path forward.

\section{Acknowledgments}
T. Linke thanks the LLNL Presidential Fellowship for providing the funding for this work, a result of J. Belof's PECASE award. This work was performed under the auspices of the U.S. Department of Energy by Lawrence Livermore National Laboratory under Contract No. DE-AC52-07NA27344.



\appendix

\section{Details of the Cu Microscopic Solver} \label{AppendixVal}
To efficiently run coupled simulations, the macroscopic method automatically calls and evaluates the microscopic domain. In general, the simulation of particles can be highly sensitive, and various factors must be considered to correctly replicate physical behavior. An example of this is the choice of a thermostat, where large fluctuations and errors can be observed if one chooses inaccurate procedures \cite{ke_effects_2022}. Therefore, the automation process comes with great responsibility towards the MD simulations to ensure reliability. A validation study of the automation is thus conducted. 

We use copper as our validation material, as it has well-documented experimental and theoretical data for a broad range of conditions, and widespread engineering applications. We used the embedded atom method (EAM) interatomic potential created by Mishin et al. \cite{mishin_structural_2001}. In the EAM, the total energy of a system is described as
\begin{equation} \label{eq:Etot}
    E_{\text{tot}} = \frac{1}{2} \sum_{i,j} V(r_{ij}) + \sum_{i} F(\bar{\rho}_i),
\end{equation}
where $V(r_{ij})$ is the pair potential determined by the distance $r_{ij}$ between atoms i and j and F is the embedding energy of site i. The embedding energy describes the effect of an overlaying electron density $\bar{\rho}$ caused by all other atoms. While this form offers a physically intuitive explanation of the potential, in practice the EAM treats $V$ and $F$ as fitting functions according to, e.g. first principles MD data. The challenge lies in the underlying ab initio data, which is often obtained at equilibrium conditions, and thus limits the applicability of the fitted EAM potential. Nonetheless, Mishin’s EAM potential has been widely adopted in the literature. Sadigh et al. \cite{sadigh_metastablesolid_2021} showed that, while it was fit specifically to ambient and near-ambient conditions, it reasonably captures atomic interaction in a generic close-packed configuration under compression and analyzed metastable phases during non-equilibrium solidification. At the melt temperature, they found a stable face-centered cubic (fcc) phase for $P<71.6$ GPa and a stable body-centered cubic (bcc) phase at $P>85$ GPa. They also observed a stable hexagonal close-packed (hcp) phase in the intervals of $71.6<P<85$ GPa. While the bcc and fcc phases were later confirmed experimentally by Sims et al. \cite{sims_experimental_2022}, the hcp phase was not found. It is concluded that it is likely an artifact of the EAM potential. Thus, care must be taken when observing the corresponding pressure ranges. 

\begin{figure}
\centering
\includegraphics[width=0.7\textwidth]{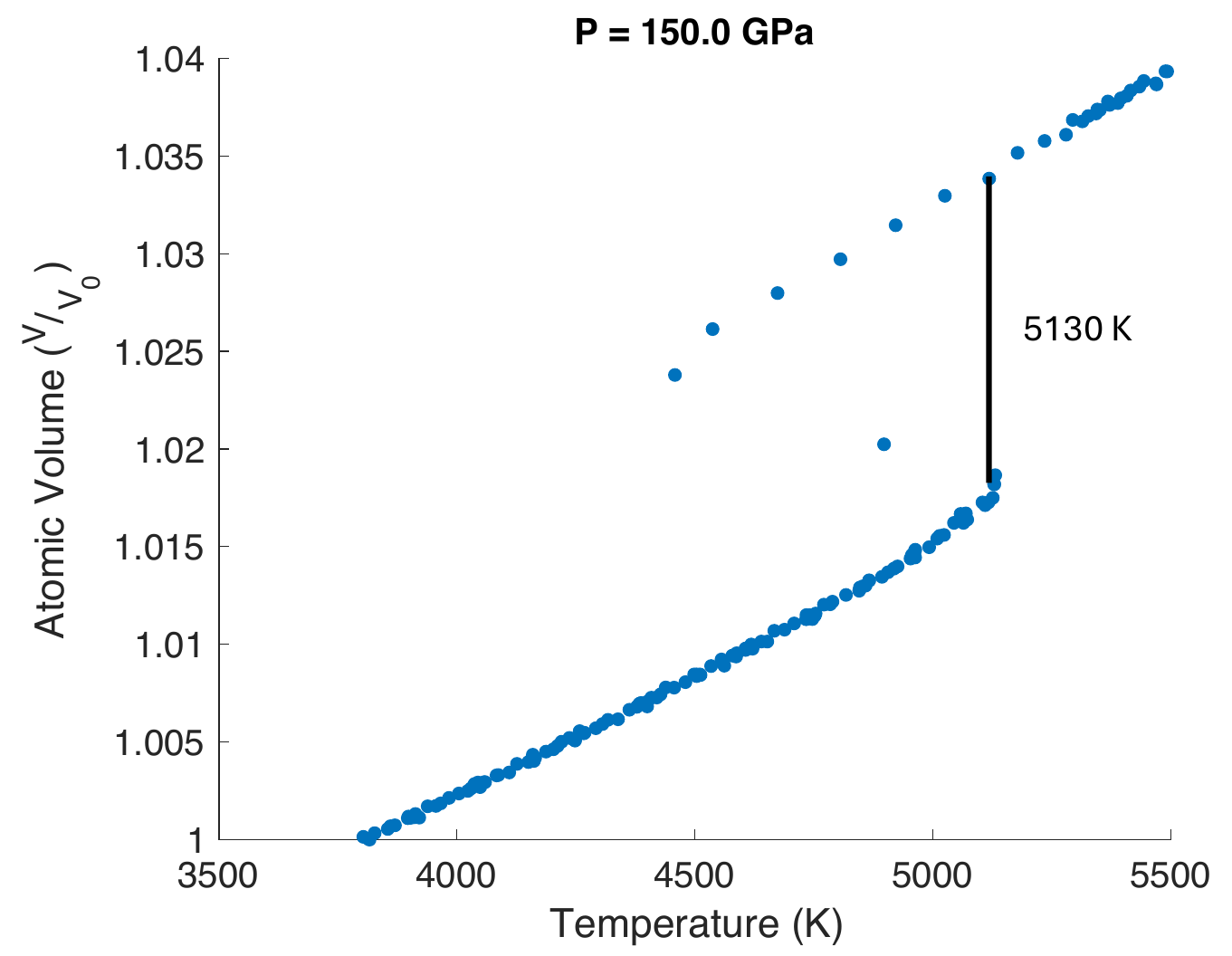}
\caption{The Heat Until Melt method increases temperature at isobaric conditions to obtain the melting point. Using the HUM method, we found that a discontinuity in the atomic volume occurs at a temperature of approximately 5130 K, indicating that a phase transition likely occurs at this temperature.}
\label{fig:HUM}
\end{figure}

Extreme conditions involve large deviations in both pressure and temperature, and our validation study must span these conditions. The calculation of the melting curve of copper using the microscopic solver employed in our coupling framework ensures that a wide range of temperatures and pressures are tested, and supplies a quantitative measure to validate against. We obtain the melting curve using the Heat Until Melt (HUM) method. HUM gradually increases the temperature of a system until it transitions from a solid to a liquid phase. A barostat is applied to maintain a constant pressure through the process. The melt temperature is identified when the atomic structure loses its crystalline order, indicating a phase transition. Figure \ref{fig:HUM} shows the typical increase seen in atomic volume at the melting point for a pressure of 150 GPa.

\begin{figure}
\centering
\includegraphics[width=0.7\textwidth]{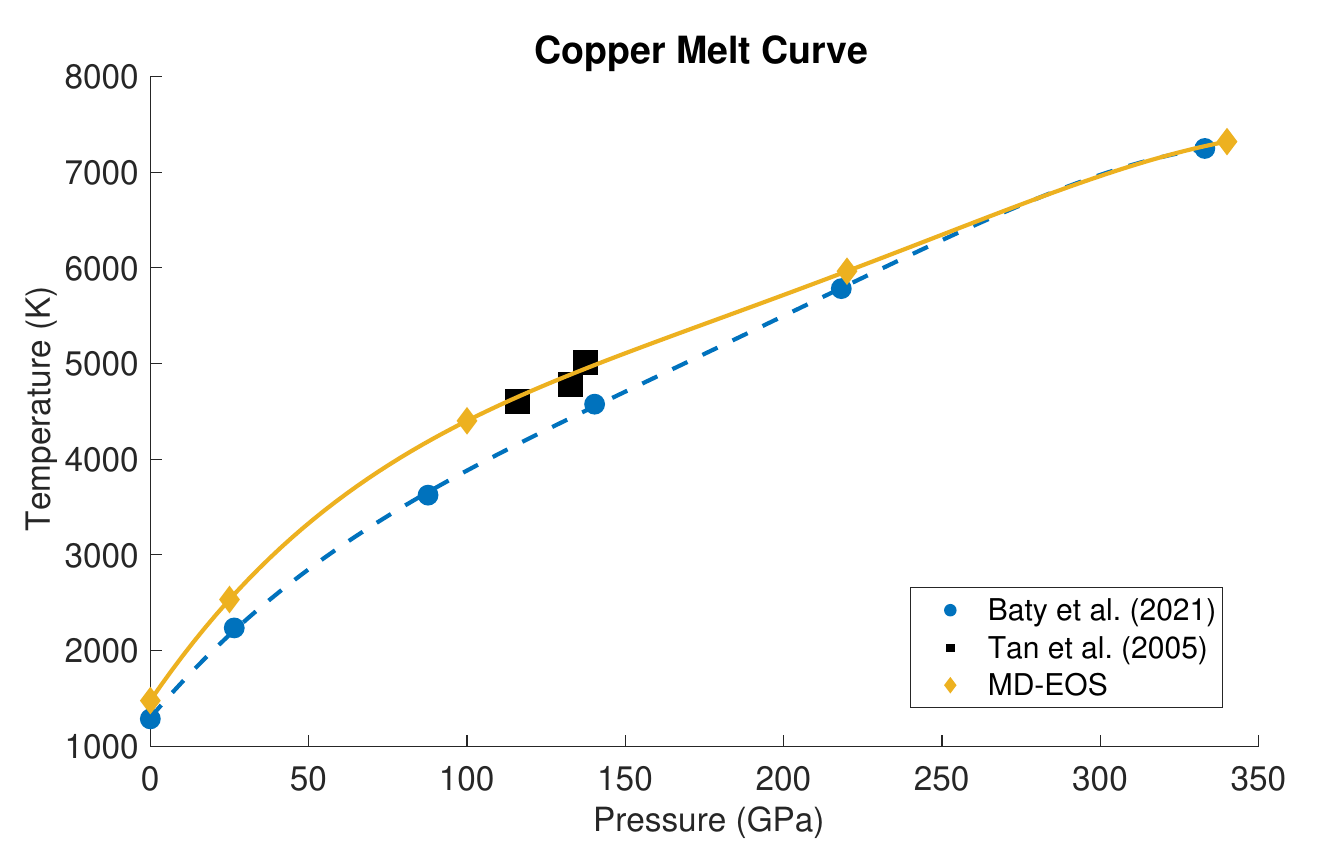}
\caption{The melting curve of copper at various pressures is compared to experimental data (black) and first principles MD simulations (blue). General agreement is observed with slight deviations at lower pressures.}
\label{fig:CuMelt}
\end{figure}

We compare the obtained melting curve with experimental data \cite{tan_method_2005} and first-principles MD calculations \cite{baty_ab_2021}. Figure \ref{fig:CuMelt} shows that the general microscopic framework presented in this work is able to obtain general agreement with prior studies of copper. The quantum molecular dynamics (QMD) data of Baty et al. (blue) shows only slight deviations from our results at pressures below 50 GPa. As pressure increases further, significant differences are observed. However, the experimental data of Tan et al. (black) in this domain is matched very well. Excellent agreement is once again observed with QMD at pressures above 200 GPa, with deviations of less than 3\%. 

The agreement of our data at the low and high ends of the pressure range amounts to an initial validation of the methodology. Nonetheless, it is the range $50<P<100$ GPa that undermines the complete validity. However, the same range was noted as problematic in the findings of Sims et al., discussed previously. This match indicates that it is not the method, but rather the interatomic potential at use that is the cause of the discrepancies. In addition, it is generally observed that our method tends to overestimate, rather than underestimate, the melting temperature. This is a known artifact of the HUM method, since it does not introduce defects into the crystal model \cite{wu_melting_2011}. In conclusion, we do not observe any computationally rooted errors within our method and deem our microscopic solver routine as trustworthy.

\section{Suitability of the Tabulated Equation of State} \label{AppendixSuit}

To evaluate the differences between an atomistic equation of state and its tabulated counterpart, it is important to establish a direct, one-to-one comparison. To this end, we derived a tabulated EOS directly from the interatomic potential for copper developed by Mishin et al. \cite{mishin_structural_2001}. This potential is widely used for MD simulations of copper and serves as a well-established reference. By constructing a table based on the same potential, we attempt to minimize any discrepancies arising strictly from the choice of MD vs. tabular EOS descriptions.

The table is constructed using the free energy of the form
\begin{equation}
    F(\rho,T) = F_\textrm{cold}(\rho,T) + F_\textrm{ion}(\rho,T).
\end{equation}
For each phase, the free energy components were fit based on data obtained from molecular dynamics simulations using the Mishin potential. The MD simulations focused specifically on accurately capturing the melt line of the potential. The separate phases were then combined into a joint table. The resulting table was integrated into our continuum code using the LLNL-developed \textit{LEOS} library. 

The tabulated EOS is designed to be an accurate representation of the Mishin-Cu EOS within a density range of $\rho = [5\ 25] \frac{\textrm{g}}{\textrm{cm}^3}$ and temperature $T=[2000\ 5000]$ K. We evaluate our atomistic EOS within the same sample space and obtain energy $e$ and pressure $p$ values for 77 different $\rho-T$ state pairs. We perform the MD simulations using a system of 4000 atoms, which showed agreement with systems of 108000 atoms across all conditions, rendering the system size sufficiently large. The simulations ran for 6,000 timesteps, the last 2000 of which were used to sample the macroscopic properties.

\begin{figure}
\centering
\includegraphics[width=0.9\textwidth]{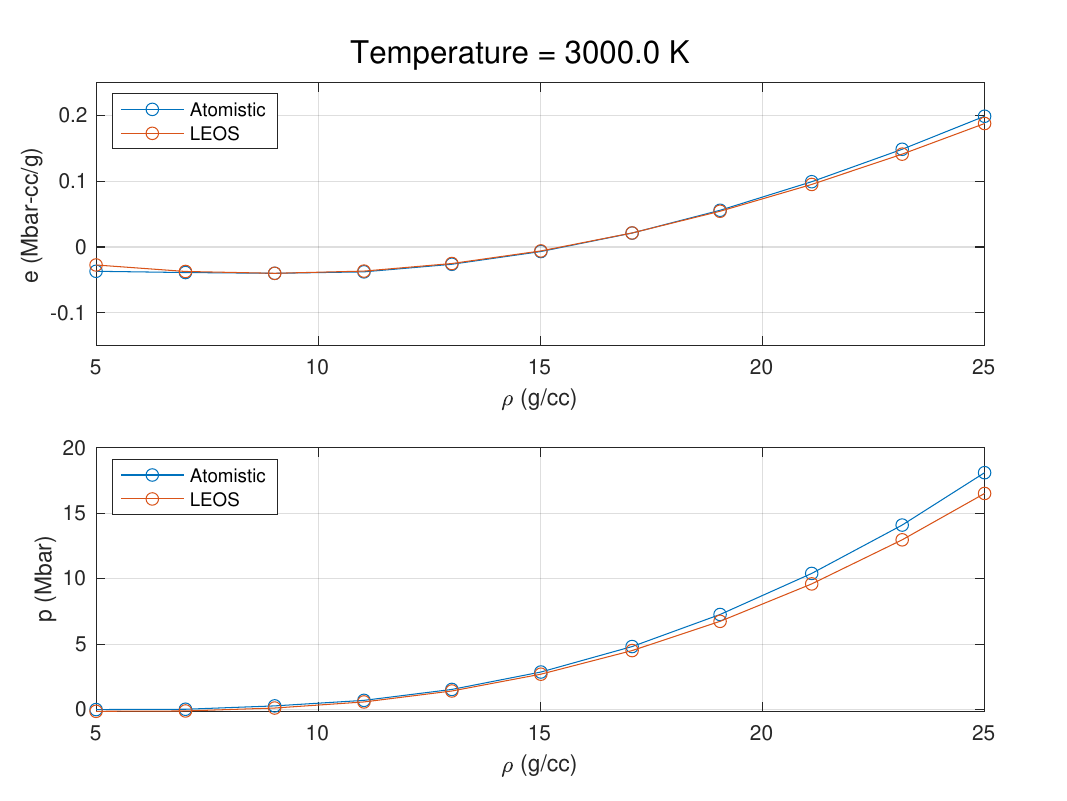}
\caption{The atomistic (blue) and tabulated (orange) equation of state models are compared across varying densities at a constant temperature $T=3000$ K.}
\label{fig:T3000}
\end{figure}

Figure \ref{fig:T3000} shows the comparison between 11 such pairs sharing the same temperature $T=3000$ K. Both energy and pressure results are compared. Generally, excellent agreement is observed between the atomistic EOS and the tabulated EOS for energy and pressure. Especially in the range $\rho = [7\ 19] \frac{\textrm{g}}{\textrm{cm}^3}$, the two methods deliver indistinguishable results. For the low density of $\rho=5 \frac{\textrm{g}}{\textrm{cm}^3}$, initial deviations between the models are observed in the energy, but not in the pressure. For the higher compressions of $\rho > 17 \frac{\textrm{g}}{\textrm{cm}^3}$, the two methods begin to gradually diverge in both energy and pressure results. The maximum absolute deviation between both models is observed at $\rho = 25 \frac{\textrm{g}}{\textrm{cm}^3}$ for energy as well as pressure, showing a relative error of $5.9\%$ and $9.6\%$, respectively.

\begin{figure}
\centering
\includegraphics[width=0.9\textwidth]{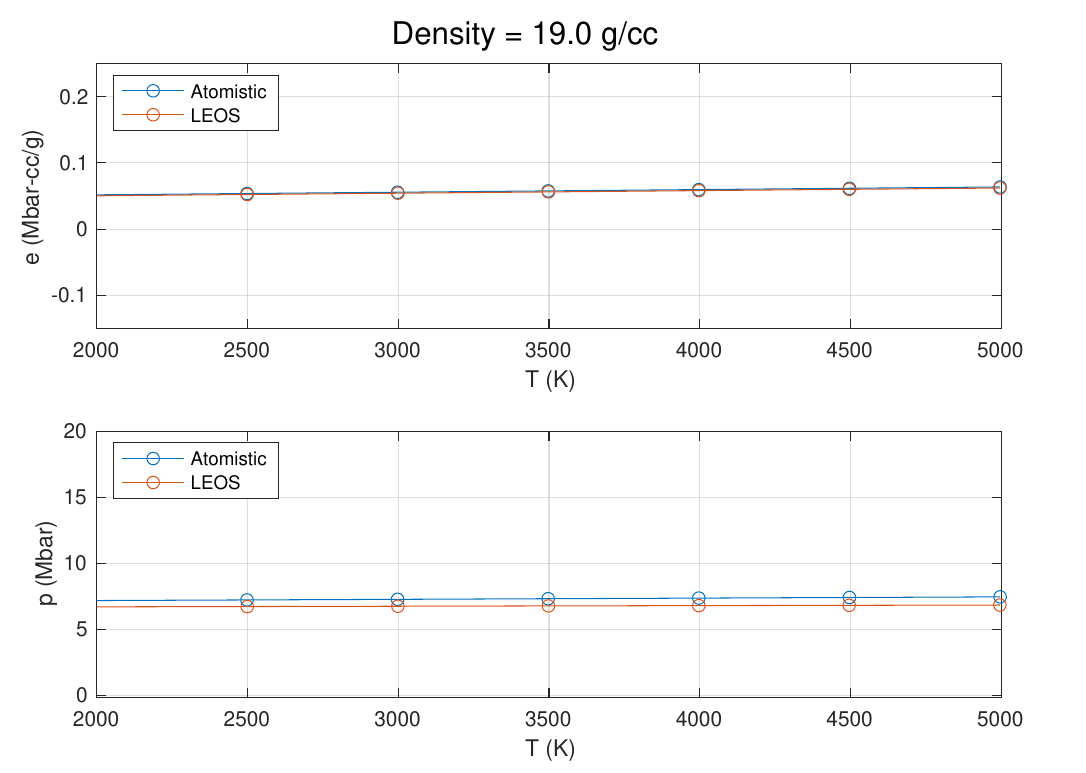}
\caption{The atomistic (blue) and tabulated (orange) equation of state models are compared across varying temperatures at a constant density $\rho=19 \frac{\textrm{g}}{\textrm{cm}^3}$.}
\label{fig:rho19}
\end{figure}

A similar comparison can be conducted for varying temperatures under constant densities. Figure \ref{fig:rho19} presents a detailed comparison between 7 data pairs at a fixed density of $\rho = 19 \frac{\textrm{g}}{\textrm{cm}^3}$, evaluating both energy and pressure as functions of temperature. Overall, the results demonstrate very good agreement between the atomistic equation of state and the tabulated EOS derived from the same interatomic potential. The two methods yield virtually identical results for energy. For the pressure results, a slight and consistent deviation is observed across the entire temperature range. Nevertheless, the deviation between the two models remains within acceptable bounds, never exceeding a relative error of $2.8\%$ in energy and $9.0\%$ in pressure.

The isotherm plot of Figure \ref{fig:T3000} shows increasing deviations between the atomistic and tabulated EOS at the low and high density domain boundaries. In contrast, the isochore plot of Figure \ref{fig:rho19} shows no significant deviations with respect to changing temperatures. This trend is observed across all sample space slices; varying densities have a much stronger effect on the accuracy of the models than varying temperatures. This behavior can be attributed to the fact that density directly influences atomic interactions. The configuration of the system is strongly coupled to the potential energy as well as the pressure. In contrast, temperature primarily alters the kinetic energy, and does not significantly modify the atomic configuration. As a result, the models remain more robust across a wide temperature range, while their accuracy becomes more responsive to changes in density. This is particularly noticeably in high and low density regimes of strong compression or rarefaction.

\begin{figure}
\centering
\includegraphics[width=0.7\textwidth]{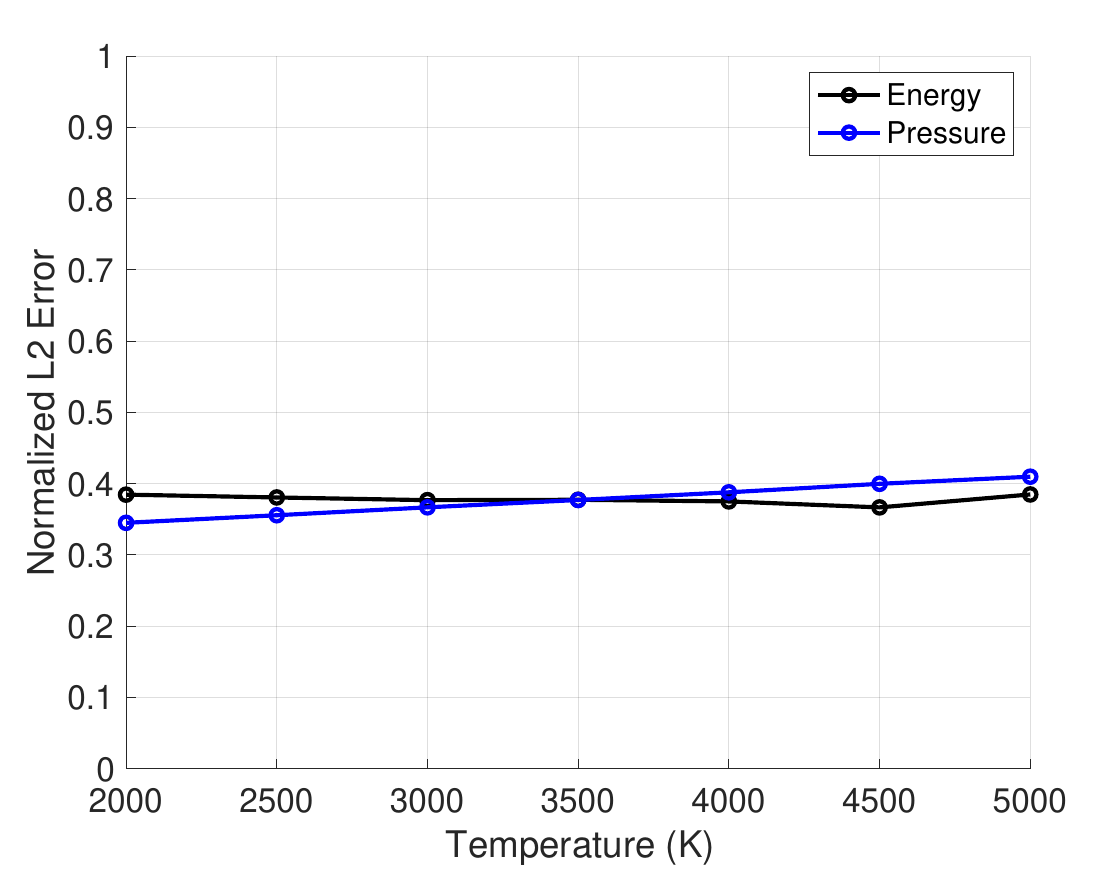}
\caption{The L2 error norm in energy (black) and pressure (blue) between the atomistic and tabulated EOS across constant temperatures show no bias towards a temperature regime.}
\label{fig:L2Temp}
\end{figure}

To determine the largest contributors to deviations between the two models, we investigate the difference between pressure and energy across all 77 sample points. For each constant temperature, we obtain the L2 error norm across the 11 densities. The resulting values for the 7 temperatures are normalized to assess the contribution each constant temperature has to be overall deviations. Figure \ref{fig:L2Temp} presents the normalized error contribution. The evenly distributed error contributions in both energy and pressure confirm the initial observation above: the atomistic and tabulated EOS are very insensitive to temperature changes. Fig. \ref{fig:L2Temp} clearly shows that errors made across densities are very comparable across all temperature sets.

\begin{figure}
\centering
\includegraphics[width=0.7\textwidth]{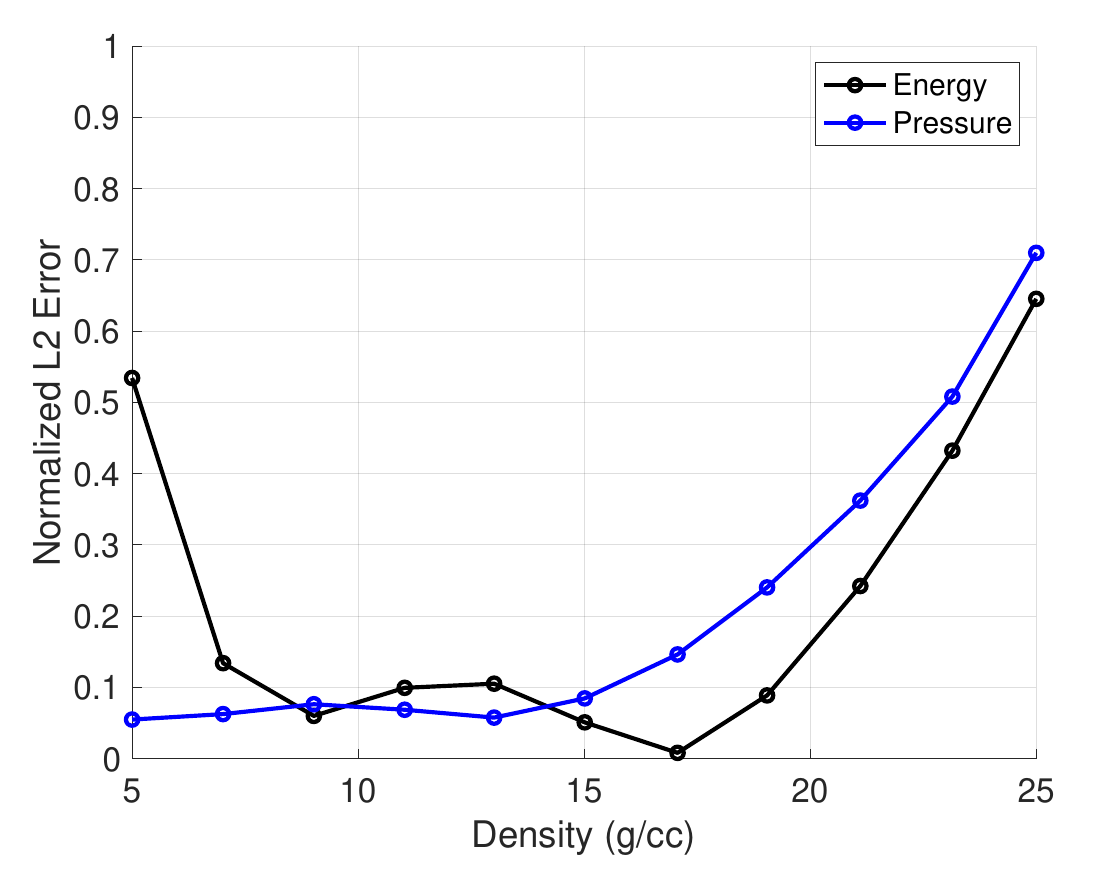}
\caption{The L2 error norm in energy (black) and pressure (blue) between the atomistic and tabulated EOS across constant densities show that the low and high density regimes contribute largely to the overall deviations.}
\label{fig:L2Density}
\end{figure}

Similarly, we compare the contribution to the overall deviations made across densities. Figure \ref{fig:L2Density} shows the normalized L2 norm obtained for constant densities, sampled across the whole temperature range for each density. The plot clearly shows that the discrepancies between the two methods occur primarily in the high density regime. For $\rho \geq 19 \frac{\textrm{g}}{\textrm{cm}^3}$, the contributions dramatically increase in both energy and pressure. The contributions of the high density regime are more pronounced in the pressure; the energy also shows significant contributions occurring at low densities. Since the pressure in these low density regimes is very close to zero, no major contributions are expected to show.

In general, this analysis shows very good agreement between the tabulated and atomistic EOS, with relative errors staying below $10\%$. Figure \ref{fig:L2Density} strongly supports that the two methods are suitable for a direct comparison of performance when integrated into a continuum code, especially in the mid-density regime. The deviations at the boundaries of the density ranges result from limitations of both methods: tabulated equations of state are constructed from discrete data points over a finite range of thermodynamic conditions, and their accuracy depends heavily on the density and temperature coverage and resolution of the original dataset, while the atomistic EOS can only guarantee accurate results throughout the validity of its interatomic potential. Near the limits of its range of validity, the interpolation/extrapolation becomes less constrained by the data, which may lead to errors. Consequently, discrepancies between the two approaches are most pronounced near the edges of their range of validity.

\section{Optimization of the Continuum-Atomistic Simulation} \label{AppendixOpt}

We present the computational optimization of both the continuum as well as the atomistic simulation. The investigation in this section is by no means exhaustive, but serves as a primary guide to reduce runtime and/or computational resources needed.

\subsection{Continuum Solver}

For the continuum-atomistic framework, the number of microscopic simulations should be minimized without a reduction in accuracy. This is achieved by minimizing the number of quadrature point evaluations in the continuum solver, which are directly determined by, but not exclusive to, the number of elements in the domain and the number of total timesteps. For the entirety of this work, we elected to obtain an EOS measurement from MD for \textit{every} EOS call. That is, for every element, and for every timestep, we created an independent microscopic instantiation. This decision was made to examine the computational feasibility of the approach in the extreme case of maximal calls to the MD, as well as to avoid any compromises in accuracy that might have resulted from limiting the number of these calls. Nonetheless, we worked to minimize the overall number of EOS evaluations required on the continuum solver side without a sacrifice in accuracy. That analysis is shown here.

\begin{figure}
\centering
\includegraphics[width=0.9\textwidth]{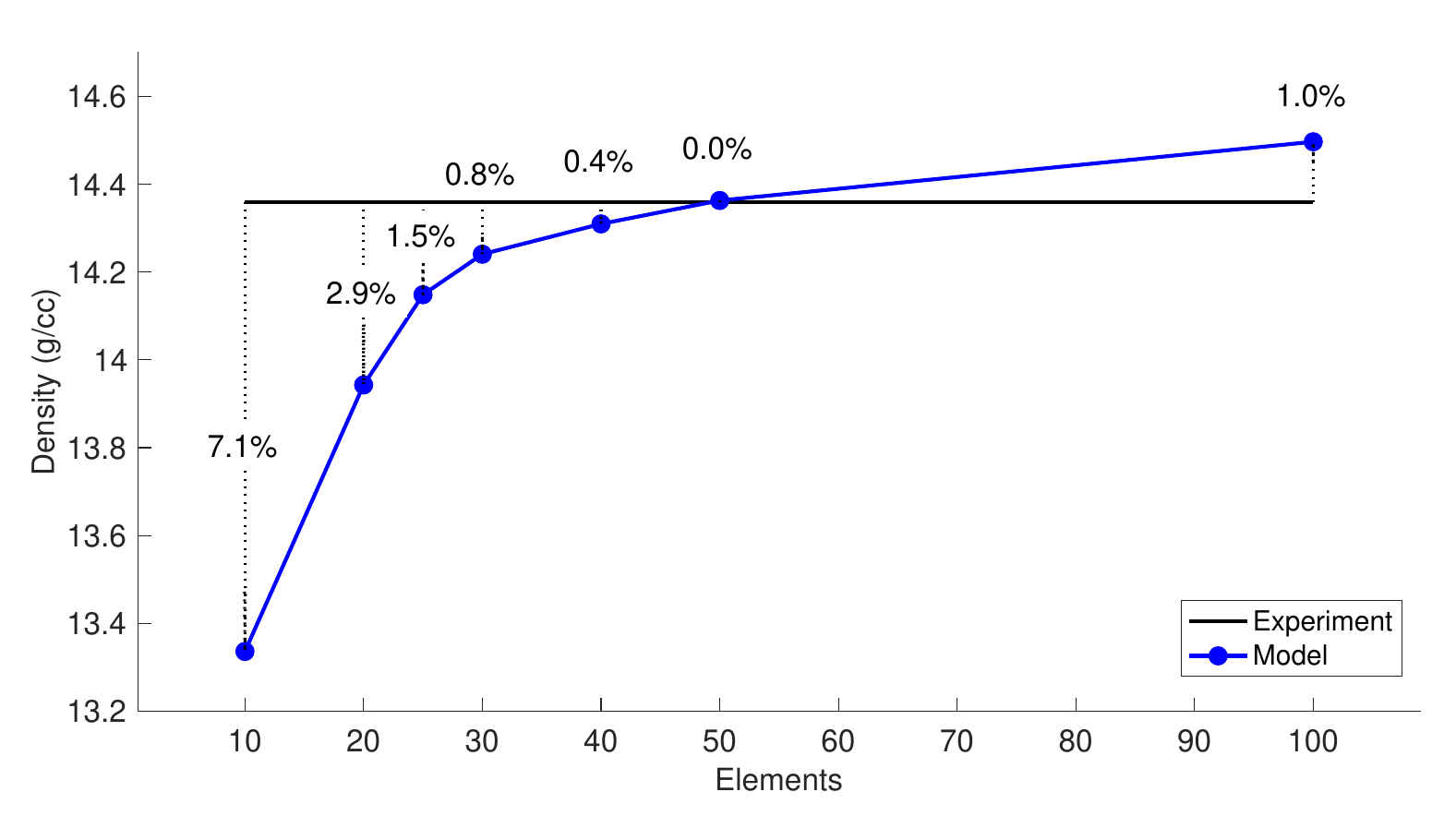}
\caption{Relative error of the density w.r.t. the number of elements used in the model}
\label{fig:DensError}
\end{figure}

To identify the minimum number of elements, a spatial discretization study is conducted. Using an analytical EOS for the high velocity impact case presented in Section \ref{sec:extreme}, we alter the number of elements. A numerical convergence study had shown that the continuum solver becomes unstable and fails when using more than 100 elements to discretize the domain. Due to this lack of a benchmark, we compare with the experimental data \cite{mitchell_shock_1981} to generally assess the difference in accuracy among discretization choices. Figure \ref{fig:DensError} presents the density of the model with respect to the number of elements used to discretize the quasi-1D domain. When using very few elements, we observe large deviations compared to the use of e.g. 100 elements. For as few as 10 elements, we observe a relative deviation of over $7\%$ between the model and the experimental data. The discrepancy  dramatically decreases when doubling the element count, reaching a relative deviation of $2.9\%$. This trend continues, until only slight changes in density occur. For instance, increasing the number of elements from 30 to 40 yields a change in density of $0.4\%$ compared to the experimental data. Interestingly, a nearly perfect agreement with the experimental data can be achieved using 50 elements. Further increases in element count causes the density to increase further, and to deviate from the experimental runs. Figure \ref{fig:DensError} shows that the converging density behavior does not continue past 50 elements. We postulate that this may be caused by very large aspect ratios in the elements, which can lead to numerical issues. Beyond 100 elements, the solver fails due to numerical instabilities.

We repeat the analysis for pressure. Again, significant deviations are observed when considering very few elements. Continuous improvement in the relative deviation is seen among increasing number of elements. For instance, an increase in number of elements from 30 to 40 exhibits a change in relative deviation of only $1.2\%$. While another improvement in relative deviation is achieved between 40 and 50 elements, we again observe a diverging behavior when increasing the number of elements to 100. To provide a baseline, we also compare the pressure results to the experimental data and observe a relative error of e.g. $3.8\%$ for 50 elements. Contrary to the density study, using more than 50 elements yields better agreement with experimental data; 100 elements show a very small relative deviation of $0.3\%$.
\begin{figure}
\centering
\includegraphics[width=0.9\textwidth]{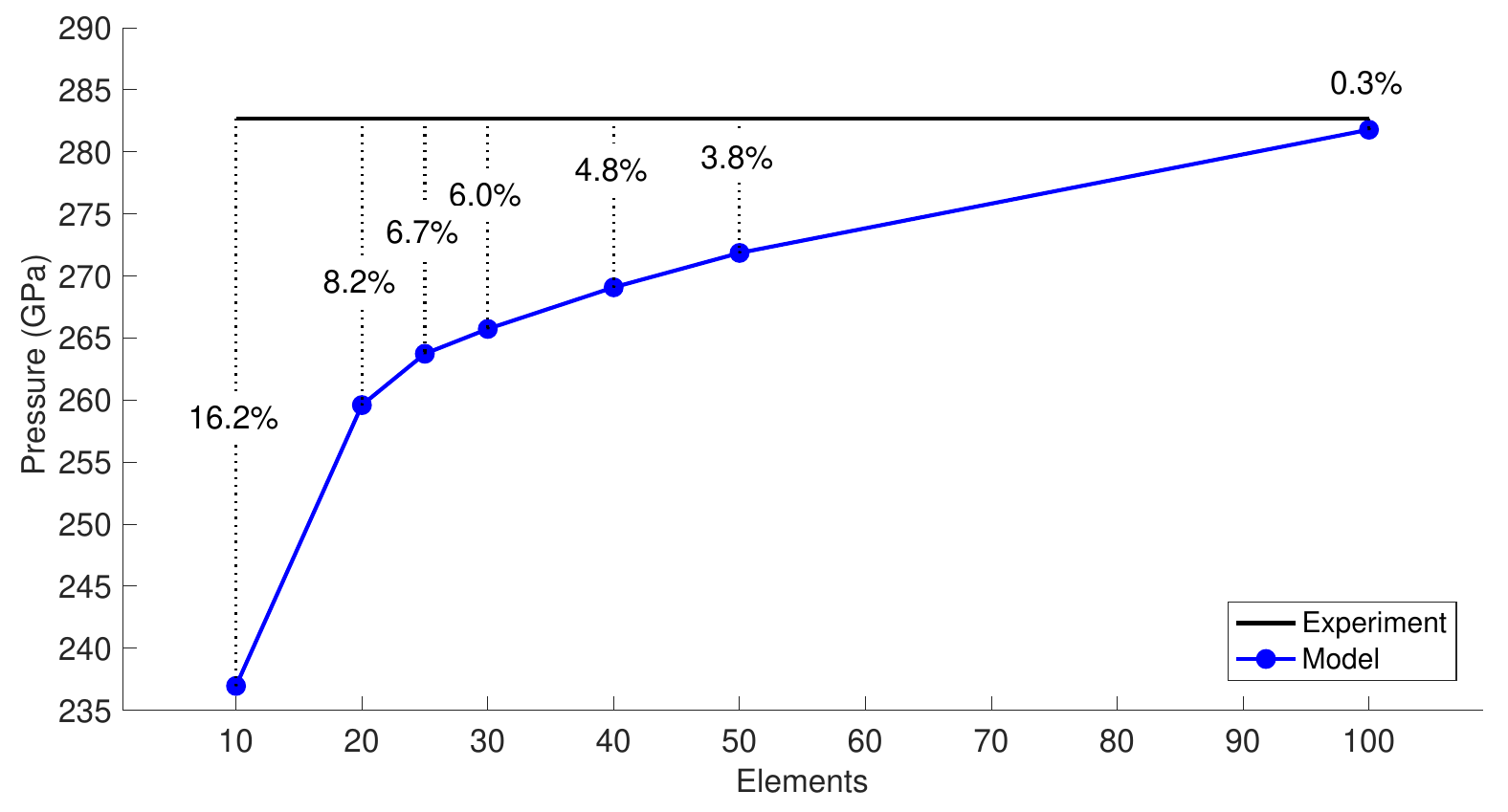}
\caption{Relative error of the pressure w.r.t. the number of elements used in the model}
\label{fig:Perror}
\end{figure}

The density as well as pressure studies confirm that using a larger number of elements improves the accuracy of the simulation; with an increased number of elements, the relative deviation to a benchmark decreases. This observation holds for up to 50 elements, beyond which numerical instabilities are assumed to cause discrepancies from a physically accurate solution. The optimal choice in element count should therefore be chosen such that the relative deviation to neighboring counts, both slightly more and slightly fewer elements, is sufficiently small. As this study aims to isolate the impact of the material model on accuracy, we choose the highest number of elements before the onset of numerical issues. In agreement with the analysis above, we conduct all simulations with a spatial discretization of 50 elements.

Interestingly, Figures \ref{fig:DensError} and \ref{fig:Perror} show different relative errors from the experimental data per element count. While a near-perfect match in density is achieved using 50 elements, the corresponding pressure error reads $3.8\%$. Similarly, the relative error in pressure for 10 elements is more than doubled compared to the density error. And for 100 elements, a very close match in pressure is observed ($0.3\%$), while the density error increases to $1.0\%$. To explain this inconsistency, we consider the roles of both density and pressure within the continuum simulation: density is explicitly evolved by the macroscopic conservation equations, whereas pressure is computed indirectly through the equation of state. From this perspective, density serves as the primary variable governing the physical evolution, while pressure is a derived quantity. Consequently, a match in density takes precedence when determining the accuracy of the continuum solver, and our choice of 50 elements is confirmed to agree well with experiments.

The final number of required elements is obtained after an initial test run with the discretized domain. As shock-driven flows are characterized by the pre- and post-shock domains, we observe which cells remain unchanged (pre-shock) when the Hugoniot states are obtained. This allows us to simply shorten the total domain to the minimal physical length required, without changing the pre-determined geometry of each element. Based on this rationale, we conclude that the total number of elements required to capture the problem most accurately is 28 elements. 

Similarly, the time step size $\Delta t$ can be maximized to reduce the number of total timesteps required. We repeat the above analysis, altering the time step $\Delta t$ and observing the change in density and pressure. The stability of the explicit solver worsens within a very small window of $\Delta t$, which has negligible effects on density and pressure readings ($<0.01\%$). The choice in timestep is therefore primarily determined by numerical stability of the matrix solver. We select a timestep of $\Delta t = 0.23$ ns.

\subsection{Atomistic Solver}
The microscopic solver is integrated into the hydrodynamic simulation through the material model. The finite element method calls the material model at every integration point for every node, so a fast and reliable coupling is key for the practicality of our method. While the lifting and restricting operators perform algebraic calculations to set and retrieve data, the microscopic solver constitutes the majority of the workload. Efficiency in this step is crucial, and a determining factor of its performance is the number of atoms in the domain. As noted in Equation \ref{eq:numAtoms}, the choice in domain size automatically determines the number of atoms. To assist the effort of the MD simulation, we tested various system sizes to determine the minimum number of atoms necessary for accurate results.

While basic equilibrium properties may be easily obtained from very small systems, we must ensure that more complex effects can be captured as well. Therefore, we draw upon the prediction of the melting temperature in copper that was previously discussed in Appendix \ref{AppendixVal}. 

\begin{figure}
\centering
\includegraphics[width=0.8\textwidth]{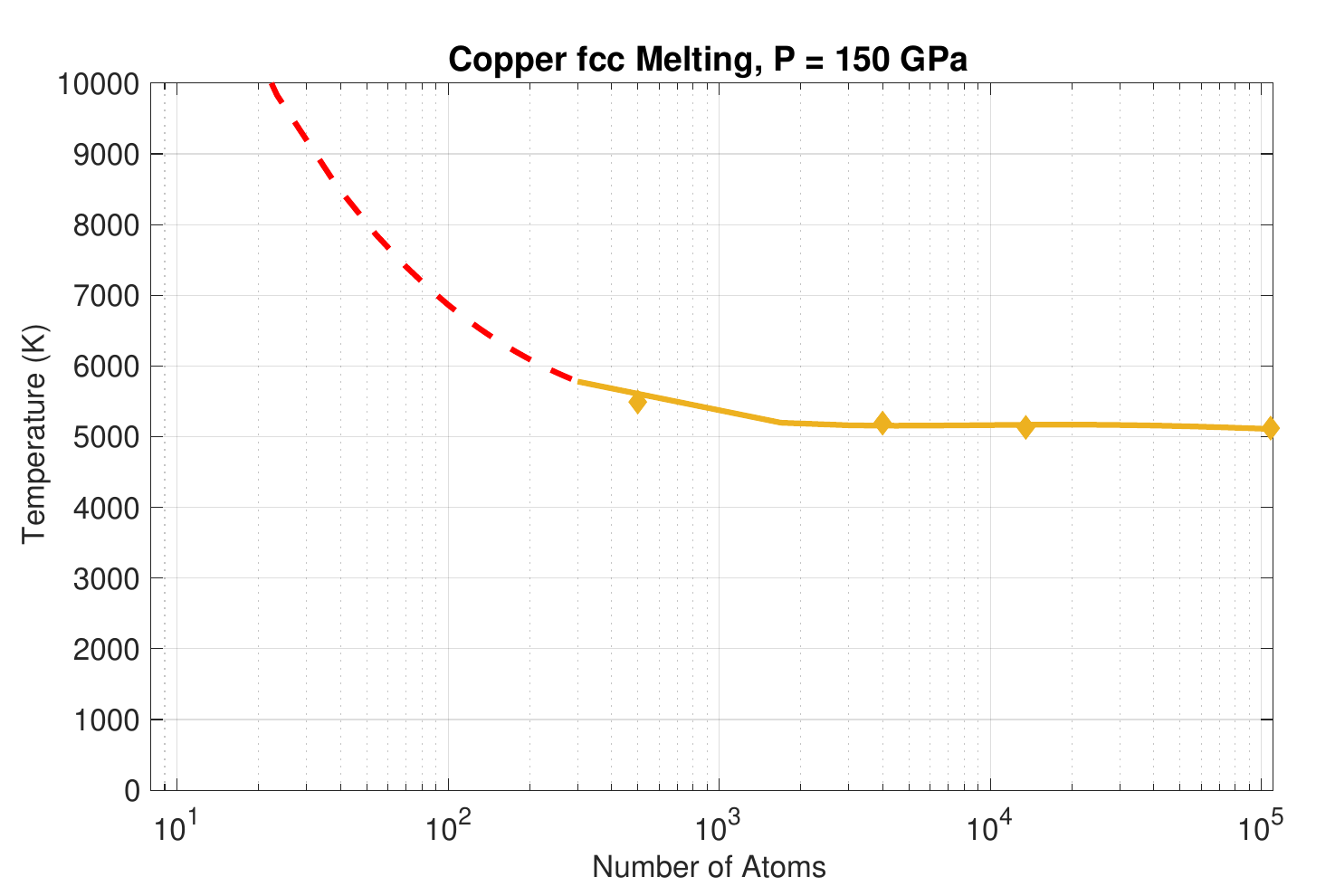}
\caption{The system size analysis shows that fewer than 400 atoms fail to produce consistent results, 500 atoms show a 7.2\% error, and systems with 4000 atoms or more achieve errors below 1\%.}
\label{fig:SystemSize}
\end{figure}

Figure \ref{fig:SystemSize} shows the melting temperature obtained at a pressure of 150 GPa for various system sizes. We observe unphysical behavior for systems smaller than 400 atoms. Strong fluctuations in the volume-temperature plots (such as Figure \ref{fig:HUM}) fail to identify a melting point, deeming such small systems useless. Acceptable deviations are first observed for systems of 500 atoms, which show a 7.2\% error. Error margins below 1\% are observed for systems of 4000 atoms or more. Equivalent findings were reported for copper in \cite{sims_experimental_2022}. The system’s ability to capture the relevant phenomena depends highly on the physical behavior that is at play, and the system size should be checked and chosen accordingly. This study used systems of 4000 atoms.

We note that to capture more complex phenomena such as phase transition kinetics, much larger systems are likely required. Such systems will carry much longer runtime durations. To enable these atomistic simulations while maintaining feasibility, innovative approaches to reduce the number of EOS evaluations should be tested in future work. A few approaches have been shown in the literature, e.g. based on active learning frameworks \cite{karra_predictive_2023}.

\bibliography{apssamp}

\end{document}